\documentclass[journal, twocolumn, comsoc]{IEEEtran}

\usepackage{amsmath}
\usepackage{amssymb}
\usepackage{amsthm}

\usepackage{newtxtext}
\usepackage{newtxmath}

\usepackage{cite}
\usepackage{hyperref}

\usepackage[pdftex]{graphicx}
\graphicspath{{./Fig/}}
\usepackage[caption=false,font=footnotesize]{subfig}

\usepackage{xcolor}
\usepackage[inline]{enumitem}
\usepackage{tikz}
\usepackage{todonotes}
\usepackage{tcolorbox}

\usepackage{array}
\usepackage{booktabs}
\usepackage{ragged2e}
\usepackage{tabularx}

\newcolumntype{L}[1]{>{\RaggedRight\arraybackslash}p{#1}}
\newcolumntype{C}[1]{>{\centering\arraybackslash}p{#1}}
\definecolor{legendUEscatterer}{HTML}{FD0103}
\definecolor{legendScattererBS}{HTML}{FD8101}
\definecolor{legendUEBD}{HTML}{39C301}
\definecolor{legendBDscatterer}{HTML}{01C3C2}
\definecolor{legendUERIS}{HTML}{C3AC01}
\definecolor{legendRISscatterer}{HTML}{FAEF01}
\definecolor{legendAngular}{HTML}{0000FF}
\definecolor{legendScatterer}{HTML}{EA01FF}
\DeclareRobustCommand{\legendarrow}[1]{%
  \tikz[baseline=-0.45ex]{\draw[#1,line width=1.4pt,-stealth] (0,0) -- (1.35em,0);}%
}
\DeclareRobustCommand{\legenddasharrow}[1]{%
  \tikz[baseline=-0.45ex]{\draw[#1,line width=1.4pt,dash pattern=on 2.2pt off 1.2pt,-stealth] (0,0) -- (1.35em,0);}%
}

\newtheorem{proposition}{Proposition}
\newtheorem{corollary}{Corollary}
\newtheorem{remark}{Remark}

\renewcommand{\Re}{\mathrm{Re}}
\renewcommand{\Im}{\mathrm{Im}}

\newcommand{\diag}{\mathrm{diag}}

\newcommand{\figswatch}[1]{\protect\raisebox{0.15ex}{\protect\textcolor[HTML]{#1}{\protect\rule{0.85em}{0.85ex}}}}

\definecolor{jeta}{HTML}{1f77b4}
\definecolor{jetb}{HTML}{ff7f0e}
\definecolor{jetc}{HTML}{2ca02c}
\definecolor{jetd}{HTML}{d62728}
\definecolor{jete}{HTML}{9467bd}
\definecolor{jetf}{HTML}{8c564b}

\begin{document}

\title{Ambient IoT Backscatter Devices as Passive Anchors for NLOS Cellular
    Positioning:\\
    Fundamental Limits}

\author{H\"{u}seyin~Yi\u{g}itler,
        Musa~Furkan~Keskin,~\IEEEmembership{Member,~IEEE,}
        Ossi~Kaltiokallio,~\IEEEmembership{Member,~IEEE,}
        and~Riku~J\"{a}ntti,~\IEEEmembership{Senior~Member,~IEEE}%
\thanks{Manuscript received July 1, 2026.
        (Corresponding author: H\"{u}seyin~Yi\u{g}itler.)}%
\thanks{H\"{u}seyin~Yi\u{g}itler and Riku~J\"{a}ntti are with the Department of
        Information and Communications Engineering, Aalto University, Espoo, Finland
        (e-mail: firstname.lastname@aalto.fi).}%
\thanks{Musa~Furkan~Keskin is with the Department of Electrical Engineering,
        Chalmers University of Technology, Gothenburg, Sweden
        (e-mail: furkan@chalmers.se).}
\thanks{Ossi~Kaltiokallio is with the Electrical Engineering Unit,
        Tampere University, Tampere, Finland
        (e-mail: ossi.kaltiokallio@tuni.fi).}}
\maketitle

\begin{abstract}

Ambient Internet-of-Things backscatter devices at known locations can act as
low-cost passive anchors by creating geometrically anchored reflected paths in
cellular networks. Unlike reconfigurable intelligent surfaces, practical
backscatter devices are independently controlled and lack a common phase
reference; their modulation signatures may be known, but their reflection gains
and residual phases are generally uncalibrated. We study how much localization
information survives this incomplete per-device calibration in uplink
non-line-of-sight (NLOS) positioning, where the direct NLOS path and the
backscatter-assisted paths share an unknown scatterer. Treating the common
channel gain, the relative backscatter response, and the residual device phases
as nuisance parameters, we derive closed-form equivalent Fisher information
matrices for calibrated, partially calibrated, and fully uncalibrated operation.
The analysis shows that unknown device phases remove carrier-phase information
from the backscatter-assisted paths, whereas joint uncertainty in the common
gain and relative response leaves the direct NLOS path with only
bandwidth-dependent delay information. The resulting position-domain bounds show
that device count alone is insufficient: the passive anchors must also observe
the common scatterer from sufficiently diverse directions. For joint
single-snapshot identification of the user equipment and scatterer, at least two
devices in two dimensions and three in three dimensions are necessary. The
results identify deployment implications for Ambient Internet-of-Things
positioning and show which calibration losses also apply to separable
subpanel-based reconfigurable-surface architectures.
\end{abstract}


\section{Introduction}
\label{sec:introduction}

Position information is an essential contextual attribute for a wide range of
mobile and Internet-of-Things (IoT) services~\cite{perera2014contextAware}.
Global Navigation Satellite System (GNSS) provides the predominant positioning
solution when satellite visibility is adequate, but their performance degrades
severely in indoor and dense urban environments. This limitation has motivated
the development of positioning methods that exploit the widely deployed cellular
infrastructure~\cite{li2026indoorPositioning}. Recent cellular generations have
introduced dedicated positioning reference signals, architectures, and
procedures~\cite{threeGpp2024ngRanPositioning,yang2022overviewThreeGpp}, enabling
geometric measurements such as time of arrival (ToA), time difference of arrival
(TDoA), and angle of arrival (AoA). These measurements are most reliably
extracted from the line-of-sight (LOS) propagation path between the base station
(BS) and the user equipment
(UE)~\cite{delPeralRosado2018surveyCellular,dwivedi2021positioningFiveG}.
Indoor radio propagation, however, is frequently dominated by non-line-of-sight
(NLOS) conditions, in which the LOS path is blocked and the received signal
consists primarily of reflected, scattered, and diffracted multipath
components~\cite{li2026indoorPositioning,abuyaghi2025positioningFiveG}.
Consequently, the measured propagation parameters no longer represent the true
geometric relationship between the BS and UE. ToA-based ranging tends to
overestimate the distance to the UE, whereas angle-based methods may infer
incorrect directions from dominant reflections and
scattering~\cite{italiano2025tutorialFiveG}. As a result, localization based
solely on conventional geometric measurements becomes challenging
and, in severe NLOS scenarios, ill-posed because reliable LOS-path information is
unavailable.

To address these limitations, recent research has explored approaches that
transform the wireless environment from an uncontrollable propagation medium into
an active participant in localization. A prominent example is the Reconfigurable
Intelligent Surface (RIS), which uses programmable metasurfaces to manipulate
propagation and create favorable radio
conditions~\cite{diRenzo2020smartRadio}. Acting as a passive
anchor at a known location, an RIS can establish a controllable reflected
path---often described as a virtual line-of-sight path---when the direct link is
blocked. Because the RIS location is known and its controllable reflection renders
this path distinguishable as a separate measurement, the same geometric
measurements can then be applied to the virtual-LOS path, restoring localization
capability under blockage~\cite{ma2023reconfigurableLocalization}.
More broadly, this demonstrates that introducing controllable, geometrically known
objects into the propagation environment can recover localization capability under
challenging NLOS conditions.

Cellular IoT systems are beginning to populate the environment with another type
of deliberately placed, known-location device that reshapes radio propagation
but, unlike the RIS, arises from a connectivity need rather than from propagation
engineering.
Under the Ambient IoT paradigm, ultra-low-power backscatter devices (BDs)
communicate without generating their own carrier waveforms, instead harvesting
energy from ambient sources and transmitting by modulating and reflecting
existing radio signals~\cite{butt2024ambientIoT,jantti2025integrationBackscatter}. A typical
link comprises a legacy cellular transmitter, one or more BDs, and a receiver
that jointly observes the direct and backscattered components. Although Ambient
IoT is motivated primarily by sustainable connectivity and massive device
deployment, a BD at a known location can also serve as a passive positioning
anchor by creating bistatic propagation paths tied to its
position~\cite{liu2016backPos,elSanhoury2025zeroEnergy}. However, unlike the
engineered anchors generally considered in RIS-assisted localization, practical
low-cost BDs are controlled independently and lack a common phase reference:
although each device's assigned modulation signature is known by design, its
reflection gain is set by the uncontrolled environment and its phase is not
commonly referenced. While backscatter-based localization has been demonstrated
in various
settings~\cite{bae2024supersight, shi2026n2los, lam2025mifly, elSanhoury2025zeroEnergy},
in this work we study what positioning information such uncalibrated passive
anchors fundamentally provide, and under what conditions Ambient IoT BDs can
support cellular positioning in challenging NLOS environments.

We consider a cellular uplink scenario in which one or more BDs at known
locations near the UE modulate the uplink sounding reference signal, so that
their backscattered components are observed at the BS through standard channel
estimation~\cite{jantti2025integrationBackscatter,liao2025ambientBackscatter}.
Unlike a conventional RIS, the BDs do not provide calibrated, controllable
reflected paths to the BS; instead, the direct NLOS and BD-assisted components
share a common propagation interaction, so their geometric parameters are
coupled under NLOS conditions. Backscatter link-budget constraints
confine the BDs to a compact neighborhood of the UE, so the UE and BDs
illuminate nearly the same region of the environment. When one propagation
interaction dominates~\cite{molisch2004genericModel}, as may occur in
corridor-to-room, street-canyon, and over-rooftop NLOS links, their paths can
share the same scatterer-to-BS segment. Under this dominant-interaction
approximation, we model the common segment
explicitly while retaining the distinct UE-to-scatterer and BD-to-scatterer
segments determined by the known BD locations. The corresponding BD-assisted
components are assumed to remain separately observable through known BD
signatures or sufficient delay/frequency separation. Although motivated by
Ambient IoT backscatter, the resulting framework also
applies to RIS-assisted localization architectures in which the individual
RIS-element (or subpanel) paths can be measured separately. This separable
geometry is illustrated in Fig.~\ref{fig:system-model}.

This problem is distinct from existing fundamental-limit analyses. RIS-assisted
localization typically assumes calibrated, controllable surface responses and
geometrically modeled
BS--RIS--UE routes~\cite{fascista2022risAidedLocalization,ma2023reconfigurableLocalization},
whereas multipath-assisted positioning relies on reflected components whose
geometry must be known or separately estimated~\cite{leitinger2015positionInformation}.
Backscatter and RFID systems can achieve high accuracy under blockage, but rely
on dedicated mmWave or near-field hardware and are largely system- and
measurement-driven~\cite{liu2016backPos,bae2024supersight,shi2026n2los,lam2025mifly}.
To the best of the authors' knowledge, the fundamental limits of
\emph{uncalibrated}, known-location passive anchors observed over a
\emph{shared-scatterer} NLOS channel---the Ambient IoT regime studied
here---remain unquantified. In this work, we address this gap by deriving
closed-form equivalent Fisher information matrices (EFIMs) under calibrated,
partially calibrated, and fully uncalibrated operation.
Our contributions are as follows:

\begin{itemize}[nosep, wide=0pt]

\item We formulate the shared-scatterer BD-assisted NLOS localization model with
the common scatterer-to-BS gain, relative BD response, and residual BD phases
treated as nuisance parameters.

\item We derive closed-form range-domain EFIMs for the resulting calibration
cases: fully calibrated operation, unknown BD phases, unknown common gain,
unknown relative BD response, joint gain-and-response uncertainty, and fully
uncalibrated operation.

\item We identify closed-form information-loss laws for each nuisance parameter.
Unknown BD phases eliminate BD-assisted carrier-phase ranging; joint
gain-and-response uncertainty couples the BD-assisted paths and leaves only
bandwidth-dependent ranging information on the direct NLOS path; and uncertainty
in the relative BD response alone preserves the direct NLOS path information
through a calibrated relative reference.

\item We derive \emph{necessary} single-snapshot conditions for joint
UE--scatterer identifiability from the rank of the position EFIM: at least two
BDs in 2D and three in 3D are required when both the UE and scatterer positions
are unknown. These counts are necessary but not sufficient because the BD
bearings to the shared scatterer must also make the geometry Jacobian full rank.

\item We map the EFIMs to UE position error bounds and show that BD count alone
does not determine accuracy. Poorly conditioned, nearly collinear deployments
leave the bound large, so angular diversity to the shared scatterer governs the
position-domain benefit and provides a concrete design criterion for Ambient IoT
NLOS positioning deployments.

\end{itemize}

The remainder of this paper is organized as follows.
Section~\ref{sec:related_work} reviews related work;
Section~\ref{sec:system-model} introduces the shared-scatterer NLOS signal model;
Section~\ref{sec:crb-results} derives the EFIMs and establishes the resulting
identifiability conditions and performance bounds;
Section~\ref{sec:numerical-results} presents numerical results; and
Section~\ref{sec:conclusion} concludes the paper.

\section{Related Work}
\label{sec:related_work}
BD-assisted positioning is closely related to multipath-assisted positioning,
which exploits multipath components (MPCs) as sensing observables and forms part
of the integrated sensing capabilities envisioned for
6G~\cite{gonzalezPrelcic2024isac}. Conventional multipath-assisted methods
exploit MPCs that arise opportunistically from the propagation environment and
whose geometry must be known or estimated~\cite{leitinger2015positionInformation}.
In contrast, each BD deliberately introduces an identifiable MPC, whose
modulation signature makes the BD-assisted component separately observable, and whose known
location provides a geometric reference, while the scatterer shared with the
direct NLOS path remains unknown.

BD-assisted positioning is also closely related to RIS-assisted localization,
since both use passive devices at known locations to introduce identifiable reflected
components that provide additional geometric information. The two architectures
nevertheless produce different observations. An RIS is a coherent aperture
whose elements reflect simultaneously, so the BS typically
observes a single beamformed component shaped by the surface phase
profile~\cite{wang2021jointBeam,wu2019intelligentReflecting}. BDs instead
modulate independently, allowing their paths to be separated by slow-time codes
at the BS. The resulting \emph{element-space} observation contains one
resolvable component per BD, rather than the \emph{beamspace} observation of a
coherent RIS. Although each BD has a weak, round-trip-limited link budget, the
spatially distributed components can provide the geometric diversity needed for
localization.

Fisher-information and Cram\'er--Rao analyses of RIS-assisted localization
provide the closest analytical precedent for our
work~\cite{fascista2022risAidedLocalization,ma2023reconfigurableLocalization}.
We adopt the same information-theoretic perspective but consider uncalibrated
BDs in a shared-scatterer NLOS regime, where the direct NLOS and BD-assisted
paths traverse a common unknown scatterer. RIS analyses with hardware
impairments account for imperfectly known surface
responses~\cite{ghaseminajm2022risAidedLocalization,ceniklioglu2022errorAnalysis},
while related NLOS studies use coded RIS tiles to obtain separable reflected
components~\cite{dardari2021localizationNlos} or multiple RISs for narrowband
SISO localization with a blocked direct
link~\cite{ettefagh2025frugalRis}. These formulations share aspects of our
setting, but assume controlled RIS responses and geometrically modeled
BS--RIS--UE routes.

Backscatter devices have themselves been used as localization anchors. Early
RFID systems such as BackPos exploit the carrier phase of backscattered tags to
reach centimetre-level accuracy, but rely on a line-of-sight, near-field link to
the reader~\cite{liu2016backPos}. More recent millimetre-wave backscatter
systems target NLOS operation directly. For example, SuperSight attains sub-centimetre NLOS
localization with mmWave backscatter tags~\cite{bae2024supersight}; N$^2$LoS
localizes a target in NLOS from a single mmWave radar and one backscatter tag,
jointly exploiting reflections from the tag and from surrounding
reflectors~\cite{shi2026n2los}; and MiFly performs six-degree-of-freedom self-localization
of a drone from a single mmWave backscatter anchor, even when the anchor is
visually occluded~\cite{lam2025mifly}. These systems demonstrate that backscatter
anchors can deliver high-accuracy positioning under blockage, which directly
motivates the present study. They achieve this, however, with dedicated mmWave
radar or near-field RFID hardware, short operating ranges, and specially designed
waveforms, and their analyses are predominantly system- and measurement-driven.

The practical feasibility of our cellular-uplink setting is supported by a
recent indoor demonstration in which a known-location zero-energy device
modulates the SRS of a nearby commercial UE and the serving BS detects its
code~\cite{elSanhoury2025zeroEnergy}. Detection within roughly a metre confirms
that nearby BDs can produce identifiable components from existing cellular
uplink signals, supporting the practical plausibility of the configuration
studied here.

\section{System Model}
\label{sec:system-model}

\begin{figure}[t!]
  \centering
  \includegraphics[width=0.65\linewidth]{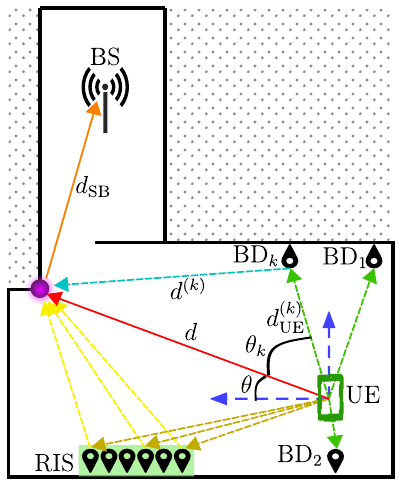}
  \caption{System-model corridor geometry for BD-assisted and separable
  RIS/subpanel-assisted NLOS positioning. Legend:
  \legendarrow{legendUEscatterer} UE--scatterer,
  \legendarrow{legendScattererBS} scatterer--BS,
  \legenddasharrow{legendUEBD} UE--BD,
  \legenddasharrow{legendBDscatterer} BD--scatterer,
  \legenddasharrow{legendUERIS} UE--RIS,
  \legenddasharrow{legendRISscatterer} RIS--scatterer, and
  \legenddasharrow{legendAngular} angular reference directions;
  \textcolor{legendScatterer}{$\bullet$} indicates the dominant scatterer.}
  \label{fig:system-model}
\end{figure}

We consider the NLOS localization scenario in
Fig.~\ref{fig:system-model}. The UE at position
$\mathbf{p}_{\mathrm{UE}}$ and a dominant scatterer at position $\mathbf{p}$
are unknown. The BS at $\mathbf{p}_{\mathrm{BS}}$ and $K$ BDs at known
positions $\{\mathbf{p}^{(k)}\}_{k=1}^{K}$ are known. The relevant distances
are
\begin{equation}
\label{eq:distances-def}
\begin{aligned}
    &d \triangleq \|\mathbf{p} - \mathbf{p}_{\mathrm{UE}}\|, \qquad
    &d_{\mathrm{SB}} \triangleq \|\mathbf{p}_{\mathrm{BS}} - \mathbf{p}\|, \\
    &d_{\mathrm{UE}}^{(k)}
        \triangleq \|\mathbf{p}^{(k)} - \mathbf{p}_{\mathrm{UE}}\|, \qquad
    &d^{(k)} \triangleq \|\mathbf{p} - \mathbf{p}^{(k)}\|.
\end{aligned}
\end{equation}
Here $d$ is the UE-to-scatterer distance, $d_{\mathrm{UE}}^{(k)}$ is
UE-to-BD$_k$, $d^{(k)}$ is BD$_k$-to-scatterer, and $d_{\mathrm{SB}}$ is
the common scatterer-to-BS segment shared by all paths.

The model has two essential restrictions. First, the direct NLOS component and
the BD-assisted components are assumed to be dominated by the same propagation
interaction point $\mathbf{p}$, as in geometry-based macro/microcell channel
models~\cite{molisch2004genericModel,molisch2003geometryDirectional}. Thus the
scatterer-to-BS segment is common to all modeled paths and can be absorbed into
an effective complex gain. Second, we do not use every MPC in a delay tap as
geometric information. Only the dominant component separated by the BD
signature is modeled explicitly; diffuse multipath, leakage, and weak
unresolved components are treated as part of the effective disturbance. This is
the setting expected in corridor-to-room, street-canyon, and over-rooftop NLOS
links where a compact UE--BD cluster illuminates a common dominant reflector.

\subsection{Measurement Extraction}

After standard SRS pilot removal and a unitary IDFT, the received delay-domain
sample in SRS burst $s$ is
\begin{equation}
\label{eq:delay-domain-channel}
    y[m,s]=\sqrt{\mathcal{P}_s}\,h[m,s]+w_t[m,s],
\end{equation}
where $m$ is the delay-bin index, $s=0,\ldots,N_s-1$ is the burst index,
$\mathcal{P}_s$ is the per-tone SRS power, and $h[m,s]$ is the sampled
delay-domain channel coefficient in bin $m$ during burst $s$. The channel is
constant within each burst but may vary across bursts because of BD switching,
and $w_t[m,s]\sim\mathcal{CN}(0,\sigma_w^2)$. The delay-bin spacing is
$T_s=1/(N\Delta_f)$, where $N$ is the IDFT size and $\Delta_f$ is the
subcarrier spacing. Appendix~\ref{appendix:measurement-extraction} details the
OFDM/SRS processing and the physical channel coefficients underlying
\eqref{eq:delay-domain-channel}.

The slow-time channel is decomposed as
\begin{equation}
\label{eq:slow-time-channel}
    h[m,s]
    =
    h_{\mathrm{env}}[m]
    +\sum_{k=1}^{K}x^{(k)}[s]h_k^{\mathrm{BD}}[m],
\end{equation}
where $h_{\mathrm{env}}[m]$ contains the quasi-static environmental
contribution, including the direct NLOS component and the structural-mode
scattering of the BDs, and $h_k^{\mathrm{BD}}[m]$ is the antenna-mode
contribution of BD$_k$. The known sequence $x^{(k)}[s]$ is constant during each
SRS burst and varies across bursts.

A unitary FFT across the slow-time index gives the delay--Doppler
representation~\cite{sturm2011waveformDesign}
\begin{equation}
\label{eq:doppler-fft}
\begin{aligned}
    Y_{\mathrm{DD}}[m,q]
    &=
    \frac{1}{\sqrt{N_s}}\sum_{s=0}^{N_s-1}
    \bigl(\sqrt{\mathcal{P}_s}h[m,s]+w_t[m,s]\bigr)
    e^{-j2\pi sq/N_s}\\
    &= \sqrt{\mathcal{P}_s}H_d[m,q]+w_d[m,q],
\end{aligned}
\end{equation}
where $H_d[m,q]$ is the slow-time DFT of $h[m,s]$, $q=0,\ldots,N_s-1$ is the
Doppler-bin index, and the unitary FFT preserves the noise variance,
$w_d[m,q]\sim\mathcal{CN}(0,\sigma_w^2)$. The
bin-to-frequency mapping is $f_D(q)=q/(N_sT_{\mathrm{rep}})$ with burst
repetition interval $T_{\mathrm{rep}}$.

Each BD applies a known slow-time switching sequence $x^{(k)}[s]$---ideally a
bin-aligned unit-amplitude tone---that places its antenna-mode contribution in
a distinct Doppler bin $q_k$. Define its unitary slow-time DFT as
\begin{equation}
\label{eq:bd-doppler-spectrum}
    X^{(k)}[q]
    \triangleq
    \frac{1}{\sqrt{N_s}}\sum_{s=0}^{N_s-1}
    x^{(k)}[s]e^{-j2\pi s q/N_s}.
\end{equation}
For $x^{(k)}[s]=e^{j2\pi s q_k/N_s}$,
$X^{(k)}[q_k]=\sqrt{N_s}$ and the other Doppler bins vanish\footnote{
The single-tone complex exponential is an
analytical abstraction; a physically realizable binary (on/off or
phase-toggling) passive switch produces a conjugate image and higher harmonics,
which place additional replicas at other Doppler bins. With a known switching
pattern these replicas fall on predictable bins and are treated as part of the
disturbance at the selected bin $q_k$, so $X^{(k)}[q_k]$ replaces $\sqrt{N_s}$
above without changing the model structure.
}. The quasi-static
environmental component is selected at $q=0$ with the same coherent factor
$\sqrt{N_s}$. These known projection factors are absorbed into the effective
gains $g_0$ and $g_k$ defined below.

Let $z_0\triangleq Y_{\mathrm{DD}}[m_0,0]$ and
$z_k\triangleq Y_{\mathrm{DD}}[m_k,q_k]$ denote the extracted direct NLOS and
BD-assisted measurements, respectively\footnote{
The indices $m_0$ and $m_k$ indicate coarse total path lengths on the delay
grid. A reflected-path delay bin alone does not determine the UE position
because the scatterer is unknown. The known delay-bin-center phase is removed
before forming the compact measurements, as detailed in
Appendix~\ref{appendix:measurement-extraction}.
}. As derived in Appendix~\ref{appendix:measurement-extraction}, their
effective complex gains are
\begin{subequations}
\label{eq:eff-gains}
\begin{align}
    g_0 &\approx \sqrt{N_s\mathcal{P}_s}\,C_0\,
    \frac{\lambda}{(4\pi)^2 d_{\mathrm{SB}}}\,
    e^{-j\frac{2\pi}{\lambda}d_{\mathrm{SB}}},\\
    g_k &\approx \sqrt{\mathcal{P}_s}\,X^{(k)}[q_k]\,
    C^{(\mathrm{a})}\widetilde{C}_k\,
    \frac{\lambda^2}{(4\pi)^3 d_{\mathrm{SB}}}\,
    e^{-j\frac{2\pi}{\lambda}d_{\mathrm{SB}}},
\end{align}
\end{subequations}
which collect the SRS amplitude, the shared scatterer-to-BS segment
$d_{\mathrm{SB}}$, the reflection/modulation and RF-chain factors ($C_0$,
$C^{(\mathrm{a})}$, $\widetilde{C}_k$), and the slow-time projection factor
($\sqrt{N_s}$ at $q=0$ and $X^{(k)}[q_k]$ at $q_k$). Define the
distance-dependent path coefficients
\begin{equation}
\label{eq:rho-def}
\begin{aligned}
    \rho_0 &\triangleq \frac{e^{-j\frac{2\pi}{\lambda}d}}{d},
    &
    \rho_k &\triangleq
    \frac{e^{-j\frac{2\pi}{\lambda}(d_{\mathrm{UE}}^{(k)}+d^{(k)})}}
         {d_{\mathrm{UE}}^{(k)}d^{(k)}} .
\end{aligned}
\end{equation}
The extracted measurements then follow the compact model
\begin{subequations}
\label{eq:meas-model}
\begin{align}
    z_0 &= g_0\rho_0+w_0,\\
    z_k &= g_k e^{j\phi_k}\rho_k+w_k,
\end{align}
\end{subequations}
where $k=1,\ldots,K$, $\phi_k$ is the residual per-BD phase offset\footnote{
Although $\phi_k$ could be absorbed into an unconstrained $g_k$, it is kept
separate because the proportional-gain reduction in
Section~\ref{sec:proportional-gain-reduction} models the gains through a common
relative coefficient.
}, and
$w_i\sim\mathcal{CN}(0,\sigma_w^2)$, $i=0,\ldots,K$, are independent under the
assumed selection of distinct orthogonal slow-time bins (or after whitening). If
an alternative delay-, frequency-, or code-based separation produces correlated
extracted noise, the covariance in~\eqref{eq:mean-covariance-fixed-g} must
instead be replaced by the corresponding post-processing covariance.
Frequency-flat carrier-phase and RF-chain offsets are absorbed into the complex
gain nuisance. We assume that any common uplink timing offset has been removed
by timing-advance synchronization, so that the extracted delays share a common
timing origin.

\subsection{Proportional-Gain Reduction}
\label{sec:proportional-gain-reduction}

The full model would contain $K{+}1$ unrelated complex gains
$\{g_0,g_1,\ldots,g_K\}$, which are not separately estimable from a single
snapshot. Each free complex gain $g_k$ absorbs the amplitude and phase of its
own path coefficient $\rho_k$, rendering that path's range parameters
$(d_{\mathrm{UE}}^{(k)},d^{(k)})$ unidentifiable, so that separating $K{+}1$
independent gains from the geometry would require multiple snapshots. To retain
single-snapshot identifiability while still exposing the calibration losses, we
share the gain across BDs through a proportional-gain reduction. The reduction
exploits structure already present in~\eqref{eq:eff-gains}, where the gains share
the scatterer-to-BS segment $d_{\mathrm{SB}}$ and its carrier phase; for
hardware-similar BDs that are close to the UE relative to the scatterer distance
and use switching sequences of equal projection magnitude, the remaining
reflection and hardware factors are well approximated by a common relative
complex coefficient,
\begin{equation}
\label{eq:gain-fixed-approx}
    g_k \approx \gamma g_0,\qquad k=1,\ldots,K,
\end{equation}
where $\gamma$ absorbs the relative antenna-mode response, wavelength scaling,
the ratio $X^{(k)}[q_k]/\sqrt{N_s}$, and any common phase or RF-chain offset.
A single $\gamma$ is therefore a calibration approximation; once the geometric
range factors are separated into $\rho_k$, the residual BD-dependent terms are
assumed common across BDs.

Define
$\boldsymbol{r}\triangleq[\rho_0,\,\rho_1 e^{j\phi_1},\ldots,\rho_K
e^{j\phi_K}]^\top$.
Writing $g\triangleq g_0$ and
$\boldsymbol{\Gamma}\triangleq\diag(1,\gamma,\ldots,\gamma)$, the joint
observation vector satisfies
\begin{equation}
\label{eq:mean-covariance-fixed-g}
    \mathbf{z} = g\,\boldsymbol{\Gamma}\boldsymbol{r}+\mathbf{w},
    \qquad
    \mathbf{w}\sim\mathcal{CN}(\mathbf{0},\sigma_w^2\mathbf{I}_{K+1}),
\end{equation}
so that $\mathbf{z}$ is Gaussian with mean $\boldsymbol{\mu}=g\boldsymbol{\Gamma}\boldsymbol{r}$ 
and covariance $\sigma_w^2\mathbf{I}_{K+1}$. 
Table~\ref{tab:parameters} summarizes the path-related quantities and nuisance
parameters in the reduced model.

\begin{table*}[t]
\caption{Path-related quantities and nuisance parameters in the reduced measurement model}
\label{tab:parameters}
\centering
\small
\setlength{\tabcolsep}{0pt}
\begin{tabularx}{\textwidth}{@{}l@{\hspace{0.8em}}L{0.21\textwidth}@{\hspace{0.8em}}X@{}}
\toprule
\textbf{Symbol} & \textbf{Related path} & \textbf{Description} \\
\midrule
$\rho_0$ & Direct NLOS path &
  Complex path coefficient of the UE-to-BS path that reflects off the
  dominant scatterer at $\mathbf{p}$ and does not involve any BD; parameterized
  by the path length $d$ and having units $\mathrm{m}^{-1}$ under
  \eqref{eq:rho-def}, while the common scatterer-to-BS segment is absorbed into
  the effective complex gain. \\
$\rho_k$ & $k$th BD-assisted NLOS path &
  Complex path coefficient/contribution of the UE-to-BS path involving the
  $k$th BD and the dominant scatterer at $\mathbf{p}$; parameterized by the path
  lengths $d_{\mathrm{UE}}^{(k)}$ and $d^{(k)}$ and having units
  $\mathrm{m}^{-2}$ under \eqref{eq:rho-def}. \\
$d$ & UE-to-scatterer segment &
  Positive real path length, in metres, of the segment between the UE and the
  dominant scatterer; couples UE and scatterer positions. \\
$d_{\mathrm{UE}}^{(k)}$ & UE-to-BD segment &
  Positive real path length, in metres, of the segment between the UE and the
  $k$th BD; primary localization distance. \\
$d^{(k)}$ & BD-to-scatterer segment &
  Positive real path length, in metres, of the segment between the $k$th BD and
  the dominant scatterer; secondary localization distance, which may be treated
  as a position nuisance parameter. \\
$d_{\mathrm{SB}}$ & Scatterer-to-BS segment &
  Positive real path length, in metres, of the common segment between the
  unknown dominant scatterer and the BS; absorbed into the effective complex
  gain and not used as a geometric observable. \\
\midrule
$g$ & All modeled NLOS paths &
  Common complex gain (after proportional-gain reduction); under the present
  range normalization it carries units of metres. \\
$\gamma$ & BD-assisted NLOS paths &
  Complex relative BD reflection/modulation coefficient including hardware and
  aperture effects, defined through the proportional relation
  $g_k\approx\gamma g_0$; under the present range normalization it has units of
  metres. \\
$\phi_k$ & $k$th BD-assisted NLOS path &
  Real per-BD residual phase offset, in radians modulo $2\pi$. \\
\bottomrule
\end{tabularx}
\end{table*}

\subsection{BD/RIS Relation}
\label{sec:bd-ris-equivalence}

The link between BDs and RISs exploited here is an equivalence of observation
models, not of hardware. Both architectures yield the same vector
measurement~\eqref{eq:mean-covariance-fixed-g}, a superposition of reflected
components whose path lengths are fixed by known passive-device positions and
whose complex responses span the full calibration range. The equivalence holds
only when these components are separately observable. A BD imprints a frequency
or slow-time signature on the incident signal, so the BS extracts a distinct
observation $z_k$ per path; a conventional RIS instead combines its element
fields coherently into a single beamformed
component~\cite{ozdogan2020intelligentReflecting} and reaches the vector form
$[z_1,\ldots,z_K]^\top$ only when its elements or subpanels are distinguished by
time/frequency coding, switching, or an equivalent mechanism.

For such separable architectures, the EFIMs, identifiability conditions, and
calibration regimes of Section~\ref{sec:crb-results} apply to element-wise or
subpanel-wise RIS localization without
re-derivation~\cite{diRenzo2020smartRadio,elzanaty2021reconfigurableLocalization}.
The BD residual phases $\{\phi_k\}$ and relative reflection coefficient $\gamma$
correspond to the element or subpanel responses relative to the panel reference,
so a calibrated RIS maps to the known-$\boldsymbol{\phi}$ regime, while phase
quantization, synchronization offsets, and element-level errors move it toward
the unknown-$\boldsymbol{\phi}$ regimes. Only the geometry differs, since
distributed BDs provide diverse reference positions whereas RIS elements are
confined to a surface with limited angular diversity. The interpretation thus
transfers our calibration and geometric conclusions to separable deployments,
not to identical hardware or scattering.

\section{Cram\'er--Rao Bound Analysis}
\label{sec:crb-results}

The compact measurement model in~\eqref{eq:mean-covariance-fixed-g} defines the
statistical link between the extracted labelled paths and the unknown geometry.
In this section, we use the Cram\'er--Rao bound (CRB)~\cite[ch.~3]{kay1993fundamentalsStatistical} to
establish fundamental lower bounds on the accuracy of any unbiased estimator
under this model. For an unbiased estimator $\hat{\boldsymbol{\eta}}$ of a
deterministic parameter vector
$\boldsymbol{\eta}$,
$\mathrm{cov}(\hat{\boldsymbol{\eta}})\succeq
\boldsymbol{\mathcal{I}}^{-1}(\boldsymbol{\eta})$, where
$\boldsymbol{\mathcal{I}}(\boldsymbol{\eta})$ is the Fisher information matrix 
and $\succeq$ denotes positive semidefinite ordering. For this Gaussian
measurement model, the covariance is fixed and all parameter dependence is carried 
by the mean $\boldsymbol{\mu}=g\boldsymbol{\Gamma}\boldsymbol{r}$. Hence, the 
FIM entries are~\cite[ch.~4]{vanTrees2013detectionEstimation}
\begin{equation}
\label{eq:fim-formula}
    [\boldsymbol{\mathcal{I}}(\boldsymbol{\eta})]_{i,j}
    =
    \frac{2}{\sigma_w^2}
    \Re\!\left\{
    \frac{\partial\boldsymbol{\mu}^{H}}{\partial\eta_i}
    \frac{\partial\boldsymbol{\mu}}{\partial\eta_j}
    \right\}.
\end{equation}

Applying the CRB requires specifying which parameters are the quantities of
interest and which are nuisance parameters. Following the terminology in
Table~\ref{tab:parameters}, the UE location affects the measurement model
through the path-length parameters
\begin{equation}
\label{eq:range-parameter-vector}
\boldsymbol{\eta}_r =
[d,\,d_{\mathrm{UE}}^{(1)},d^{(1)},\ldots,d_{\mathrm{UE}}^{(K)},d^{(K)}]^\top
\in\mathbb{R}^{2K+1}.
\end{equation}
These path lengths depend on the UE position, the known BD or RIS/subpanel
reference positions, and the dominant-scatterer position. We therefore first
derive estimator limits for the range-domain parameters and later map the
resulting EFIM to the UE position domain.

Working directly with the path lengths $\boldsymbol{\eta}_r$ avoids the need
to account for the shared UE/scatterer geometry at the estimation stage. These
path lengths become coupled only after they are expressed as functions of the
common position vector
$\boldsymbol{\theta}=[\mathbf{p}_{\mathrm{UE}}^\top,\mathbf{p}^\top]^\top$.
The change-of-variables Jacobian in Section~\ref{sec:range-position-mapping}
handles this dependency, so the range-domain FIM derivations treat each element
of $\boldsymbol{\eta}_r$ as an independent parameter.

The calibrated compact measurements retain the path-coefficient sensitivity of
the resolved peaks. In the waveform derivation provided in
Appendix~\ref{appendix:range-delay-fi}, this sensitivity corresponds to the
coefficient-gradient term $\partial a(\tau)/\partial\tau$, which is the same
amplitude and carrier-phase information captured by the compact-model mean
derivative. The remaining term comes from the local waveform shift
$\partial s(t-\tau)/\partial t$ and carries the finite-bandwidth delay
information associated with the selected delay locations. Because the real
cross-term between these two contributions vanishes for the centered baseband
waveform, the waveform-delay contribution can be added to the range-domain EFIM
without double-counting the compact path-coefficient information.
For any resolved path, the term
$2|\mu_i|^2\beta_B/\sigma_w^2$ derived in
Appendix~\ref{appendix:range-delay-fi}, where $\mu_i$ is the $i$-th element
of the measurement vector mean,
$\beta_B\triangleq(2\pi B_{\mathrm{rms}}/c)^2$ and $B_{\mathrm{rms}}$ is
the RMS bandwidth of the SRS waveform, must be
added.  Thus, the direct path adds this term to the $d$ entry, whereas the
$k$th BD-assisted path adds
\begin{equation}
\label{eq:delay-term}
    \boldsymbol{\mathcal{I}}_k^{(\mathrm{bw})}
    =
    \frac{2|\mu_k|^2}{\sigma_w^2}\,\beta_B\,
    \mathbf{1}_2\mathbf{1}_2^\top,
\end{equation}
to the $2\times2$ block associated with
$[d_{\mathrm{UE}}^{(k)},d^{(k)}]$, where $\mathbf{1}_2\triangleq[1,1]^\top$.

The remaining unknown model parameters in Table~\ref{tab:parameters} are
treated as nuisance parameters and collected in $\boldsymbol{\eta}_n$. Define
the full parameter vector as
$\boldsymbol{\eta}=[\boldsymbol{\eta}_r^\top,\boldsymbol{\eta}_n^\top]^\top$
and partition the FIM as
\begin{equation}
    \boldsymbol{\mathcal{I}}(\boldsymbol{\eta}) =
    \begin{bmatrix}
    \boldsymbol{\mathcal{I}}_{\eta_r\eta_r} &
    \boldsymbol{\mathcal{I}}_{\eta_r\eta_n} \\
    \boldsymbol{\mathcal{I}}_{\eta_n\eta_r} &
    \boldsymbol{\mathcal{I}}_{\eta_n\eta_n}
    \end{bmatrix}.
\end{equation}
The EFIM for the range parameters $\boldsymbol{\eta}_r$ is obtained by
eliminating the nuisance block through the Schur
complement~\cite[Sec.~0.8.5]{horn2013matrixAnalysis},
\begin{equation}
    \label{eq:EFIM-Schur}
    \boldsymbol{\mathcal{I}}(\boldsymbol{\eta}_r)
    \triangleq
    \boldsymbol{\mathcal{I}}_{\eta_r\eta_r}
    - \boldsymbol{\mathcal{I}}_{\eta_r\eta_n}
    \boldsymbol{\mathcal{I}}_{\eta_n\eta_n}^{-1}
    \boldsymbol{\mathcal{I}}_{\eta_n\eta_r}.
\end{equation}
If all non-range parameters are calibrated in advance, the nuisance block is
empty and the EFIM reduces to
$\boldsymbol{\mathcal{I}}_{\eta_r\eta_r}$.

The contents of $\boldsymbol{\eta}_n$ depend on which parameters are known to
the estimator. In particular, the common complex scalar $g$, the relative
passive-device path coefficient $\gamma$, and the residual phases
$\boldsymbol{\phi}$ may be calibrated in advance, or they may have to be
estimated jointly with $\boldsymbol{\eta}_r$. The propositions below evaluate
\eqref{eq:EFIM-Schur} under these calibration regimes.

\subsection{Range-Domain EFIMs}
\label{sec:range-domain-efims}

Using the EFIM definition in~\eqref{eq:EFIM-Schur}, the results below use
$r_0\triangleq\rho_0$ and
$r_k\triangleq\rho_k e^{j\phi_k}$ for the $k$th BD-assisted path. Hence
$|r_k|=|\rho_k|$ even when $\phi_k$ is unknown. We also define
$\mathbf{r}_{\mathrm{BD}}\triangleq[r_1,\ldots,r_K]^\top$,
$\|\mathbf{r}_{\mathrm{BD}}\|^2\triangleq\sum_{k=1}^{K}|r_k|^2$,
and $\mathbf{v}_k\triangleq[1/d_{\mathrm{UE}}^{(k)},1/d^{(k)}]^\top$.
We denote the direct and BD-assisted per-bin measurement SNRs by
\begin{equation}
\label{eq:snr-def}
\mathrm{SNR}_0\triangleq\frac{|g|^2|r_0|^2}{\sigma_w^2},
\qquad
\mathrm{SNR}_k\triangleq
\frac{|g|^2|\gamma|^2|r_k|^2}{\sigma_w^2},
\end{equation}
for $k=1,\ldots,K$.

\begin{proposition}[Calibrated EFIM]
\label{prop:calibrated-efim}
If $g$, $\gamma$, and $\boldsymbol{\phi}$ are all known by prior calibration, 
the EFIM is block diagonal:
\begin{equation}
    \label{eq:fim-block-diagonal}
    \boldsymbol{\mathcal{I}}^{(\mathrm{cal})} = \boldsymbol{\mathcal{I}}_{\eta_r\eta_r} =
    \begin{bmatrix}
        \mathcal{I}_d & \mathbf{0}^\top \\
        \mathbf{0} &
        \operatorname{blkdiag}(\boldsymbol{\mathcal{I}}_1,\ldots,
        \boldsymbol{\mathcal{I}}_K)
    \end{bmatrix},
\end{equation}
with $\mathcal{I}_d \in \mathbb{R}$ and $\boldsymbol{\mathcal{I}}_k \in \mathbb{R}^{2\times2}$ given by
\begin{subequations}
\label{eq:fim-blocks-cal}
\begin{align}
    \mathcal{I}_d
    &= 2\mathrm{SNR}_0
    \!\left[
    \frac{1}{d^2}+\!\left(\frac{2\pi}{\lambda}\right)^{\!2}\!+\beta_B
    \right], \label{eq:I-d-cal}\\
    \boldsymbol{\mathcal{I}}_k
    &= 2\mathrm{SNR}_k
    \!\left[
    \mathbf{v}_k\mathbf{v}_k^\top
    + \!\left(\frac{2\pi}{\lambda}\right)^{\!2}\!\mathbf{1}_2\mathbf{1}_2^\top
    + \beta_B\mathbf{1}_2\mathbf{1}_2^\top
    \right]\!. \label{eq:I-k-cal}
\end{align}
\end{subequations}
The amplitude-gradient term $\mathbf{v}_k\mathbf{v}_k^\top$ acts along the
two segment lengths separately.
The carrier-phase term is
$(2\pi/\lambda)^2\mathbf{1}_2\mathbf{1}_2^\top$, and the bandwidth term is
$\beta_B\mathbf{1}_2\mathbf{1}_2^\top$ with
$\beta_B\triangleq(2\pi B_{\mathrm{rms}}/c)^2$, as derived in
Appendix~\ref{appendix:range-delay-fi}.
Both carrier-phase and bandwidth terms act along the summed-path direction.
The proof is given in Appendix~\ref{app:efim-proofs}.
\end{proposition}

\begin{proposition}[Unknown BD phase shifts]
\label{prop:unknown-phase-efim}
Let $g$ and $\gamma$ be known, and let the nuisance vector be
$\boldsymbol{\eta}_n^{(\boldsymbol{\phi})}
\triangleq\boldsymbol{\phi}=[\phi_1,\ldots,\phi_K]^\top$. With the distance
ordering in~\eqref{eq:range-parameter-vector}, the EFIM remains block diagonal:
\begin{equation}
\label{eq:efim-block-diagonal-phi}
    \boldsymbol{\mathcal{I}}^{(\boldsymbol{\phi})}
    =
    \begin{bmatrix}
        \mathcal{I}_d^{(\boldsymbol{\phi})} & \mathbf{0}^\top \\
        \mathbf{0} &
        \operatorname{blkdiag}(\boldsymbol{\mathcal{I}}_1^{(\boldsymbol{\phi})},
        \ldots,\boldsymbol{\mathcal{I}}_K^{(\boldsymbol{\phi})})
    \end{bmatrix}.
\end{equation}
The direct NLOS block is unchanged $\mathcal{I}_d^{(\boldsymbol{\phi})}
= \mathcal{I}_d$, while each
BD-assisted block $\boldsymbol{\mathcal{I}}_k^{(\boldsymbol{\phi})} \in \mathbb{R}^{2\times2}$, 
for $k=1,\ldots,K$, is
\begin{equation}
    \boldsymbol{\mathcal{I}}_k^{(\boldsymbol{\phi})}
    =
    2\mathrm{SNR}_k
    \!\left[
    \mathbf{v}_k\mathbf{v}_k^\top
    + \beta_B\mathbf{1}_2\mathbf{1}_2^\top
    \right]\!.
    \label{eq:I-k-phi}
\end{equation}
Thus the Schur complement over $\boldsymbol{\phi}$ exactly removes the
carrier-phase term
$(2\pi/\lambda)^2\mathbf{1}_2\mathbf{1}_2^\top$ from every BD-assisted block in
Proposition~\ref{prop:calibrated-efim}. The remaining information is carried by
path-loss amplitude gradients and wideband delay.
The proof is given in Appendix~\ref{app:efim-proofs}.
\end{proposition}

\begin{proposition}[Unknown gain and relative BD coefficient]
\label{prop:unknown-gain-bd-coefficient-efim}
If $g$ and $\gamma$ are unknown while $\boldsymbol{\phi}$ is known, the direct
path remains decoupled from the BD-assisted parameters, and the EFIM has the
block form
\begin{equation}
\label{eq:efim-structure-gamma}
    \boldsymbol{\mathcal{I}}^{(g,\gamma)}
    =
    \begin{bmatrix}
        \mathcal{I}_d^{(g,\gamma)} & \mathbf{0}^\top \\
        \mathbf{0} & \boldsymbol{\mathcal{I}}_{\mathrm{BD}}^{(g,\gamma)}
    \end{bmatrix},
\end{equation}
where $\boldsymbol{\mathcal{I}}_{\mathrm{BD}}^{(g,\gamma)}\in
\mathbb{R}^{2K\times2K}$, in general, is not block diagonal and contains the
$2\times2$ blocks defined below.
The direct-path block, the BD-assisted diagonal blocks for $k=1,\ldots,K$, and
the off-diagonal blocks for $k\neq\ell$ are
\begin{subequations}
\label{eq:efim-blocks-gain}
\begin{align}
    \mathcal{I}_d^{(g,\gamma)}
    &=
    2\mathrm{SNR}_0\,\beta_B ,
    \label{eq:I-d-gain}\\
    \boldsymbol{\mathcal{I}}_{k,k}^{(g,\gamma)}
    &=
    2\mathrm{SNR}_k
    \begin{aligned}[t]
    &\Bigg[
    \left(1-\frac{|r_k|^2}{\|\mathbf{r}_{\mathrm{BD}}\|^2}\right)
    \mathbf{v}_k\mathbf{v}_k^\top
    \\
    &~+
    \left(1-\frac{|r_k|^2}{\|\mathbf{r}_{\mathrm{BD}}\|^2}\right)
    \left(\frac{2\pi}{\lambda}\right)^2\mathbf{1}_2\mathbf{1}_2^\top
    \\
    &~+
    \beta_B\mathbf{1}_2\mathbf{1}_2^\top
    \Bigg],
    \end{aligned}
    \label{eq:I-kk-gain}\\
    \boldsymbol{\mathcal{I}}_{k,\ell}^{(g,\gamma)}
    &=
    -2\mathrm{SNR}_k
    \frac{|r_\ell|^2}{\|\mathbf{r}_{\mathrm{BD}}\|^2}
    \Bigg[
    \mathbf{v}_k\mathbf{v}_\ell^\top+
    \left(\frac{2\pi}{\lambda}\right)^2
    \mathbf{1}_2\mathbf{1}_2^\top
    \Bigg].
    \label{eq:I-kl-gain}
\end{align}
\end{subequations}
The direct block in~\eqref{eq:I-d-gain} shows that eliminating the common gain
removes the direct path's narrowband amplitude-gradient and carrier-phase
terms, leaving only the bandwidth contribution. For the BD-assisted paths, the
shared nuisance pair $\{g,\gamma\}$ removes the common component of the
narrowband derivative directions. This produces the attenuation factors in the
diagonal blocks~\eqref{eq:I-kk-gain} and the nonzero off-diagonal coupling
blocks~\eqref{eq:I-kl-gain}.
The proof is given in Appendix~\ref{app:efim-proofs}.
\end{proposition}

\begin{corollary}[Unknown common gain]
\label{cor:unknown-common-gain}
If $g$ is unknown while $\gamma$ and $\boldsymbol{\phi}$ are known, the EFIM has
the block form
\begin{equation}
\label{eq:efim-structure-g-only}
    \boldsymbol{\mathcal{I}}^{(g)}
    =
    \begin{bmatrix}
        \mathcal{I}_d^{(g)} &
        \boldsymbol{\mathcal{I}}_{d,\mathrm{BD}}^{(g)} \\
        \boldsymbol{\mathcal{I}}_{\mathrm{BD},d}^{(g)} &
        \boldsymbol{\mathcal{I}}_{\mathrm{BD}}^{(g)}
    \end{bmatrix},
\end{equation}
where
$\boldsymbol{\mathcal{I}}_{\mathrm{BD},d}^{(g)}
=(\boldsymbol{\mathcal{I}}_{d,\mathrm{BD}}^{(g)})^\top$ and
$\boldsymbol{\mathcal{I}}_{d,\mathrm{BD}}^{(g)}$ is composed of the
$1\times2$ blocks
$\boldsymbol{\mathcal{I}}_{d,k}^{(g)}$, $k=1,\ldots,K$. The direct block,
direct--BD blocks, BD diagonal blocks, and BD off-diagonal blocks are
\begin{subequations}
\label{eq:efim-blocks-g-only}
\begin{align}
    \mathcal{I}_d^{(g)}
    &=
    2\mathrm{SNR}_0
    \!\left[
    \left(1-\frac{|r_0|^2}{\|\boldsymbol{\Gamma}\boldsymbol{r}\|^2}\right)
    \left(\frac{1}{d^2}+\left(\frac{2\pi}{\lambda}\right)^2\right)
    +\beta_B
    \right],
    \label{eq:I-d-g-only}\\
    \boldsymbol{\mathcal{I}}_{d,k}^{(g)}
    &=
    -2\mathrm{SNR}_0
    \frac{|\gamma|^2|r_k|^2}
    {\|\boldsymbol{\Gamma}\boldsymbol{r}\|^2}
    \left[
    \frac{1}{d}\mathbf{v}_k^\top
    +\left(\frac{2\pi}{\lambda}\right)^2\mathbf{1}_2^\top
    \right].
    \label{eq:I-dk-g-only}\\
    \boldsymbol{\mathcal{I}}_{k,k}^{(g)}
    &=
    2\mathrm{SNR}_k
    \begin{aligned}[t]
    \Bigg[
    &\left(
    1-\frac{|\gamma|^2|r_k|^2}
    {\|\boldsymbol{\Gamma}\boldsymbol{r}\|^2}
    \right)
    \left(
    \mathbf{v}_k\mathbf{v}_k^\top
    +\left(\frac{2\pi}{\lambda}\right)^2
    \mathbf{1}_2\mathbf{1}_2^\top
    \right)\\
    &+\beta_B\mathbf{1}_2\mathbf{1}_2^\top
    \Bigg],
    \end{aligned}
    \label{eq:I-kk-g-only}\\
    \boldsymbol{\mathcal{I}}_{k,\ell}^{(g)}
    &=
    -2\mathrm{SNR}_k
    \frac{|\gamma|^2|r_\ell|^2}
    {\|\boldsymbol{\Gamma}\boldsymbol{r}\|^2}
    \left[
    \mathbf{v}_k\mathbf{v}_\ell^\top
    +\left(\frac{2\pi}{\lambda}\right)^2
    \mathbf{1}_2\mathbf{1}_2^\top
    \right],
    \label{eq:I-kl-g-only}
\end{align}
\end{subequations}
for $k\neq\ell$.
Thus eliminating only the common gain couples the direct NLOS range to the
BD-assisted ranges and couples the BD-assisted ranges to one another. Because
the known $\gamma$ and $\boldsymbol{\phi}$ make the direct NLOS component a
calibrated relative reference, the common-gain loss is normalized by
$\|\boldsymbol{\Gamma}\boldsymbol{r}\|^2$ rather than by
$\|\mathbf{r}_{\mathrm{BD}}\|^2$. Consequently, both the direct NLOS and
BD-assisted narrowband information are only partially removed.
The proof is given in Appendix~\ref{app:efim-proofs}.
\end{corollary}

\begin{corollary}[Unknown relative BD coefficient]
\label{cor:individual-g-gamma}
If the complex scalar $\gamma$ is unknown while $g$ and $\boldsymbol{\phi}$ are
known, the direct block is unchanged:
$
    \mathcal{I}_d^{(\gamma)} = \mathcal{I}_d.
$
The BD-assisted submatrix is identical to
$\boldsymbol{\mathcal{I}}_{\mathrm{BD}}^{(g,\gamma)}$ in
Proposition~\ref{prop:unknown-gain-bd-coefficient-efim}, with diagonal blocks
given by~\eqref{eq:I-kk-gain} and off-diagonal blocks given
by~\eqref{eq:I-kl-gain}.
Since $\gamma$ does not enter the direct path, the Schur complement over
the $\gamma$ nuisance block incurs no loss on the direct block.
\end{corollary}

\begin{corollary}[Fully unknown passive regime]
\label{cor:fully-unknown-efim}
If $g$, $\gamma$, and $\boldsymbol{\phi}$ are all unknown, the EFIM has the
same block partition and block definitions as in
Proposition~\ref{prop:unknown-gain-bd-coefficient-efim}, with blocks given by
\begin{subequations}
\label{eq:efim-blocks-full}
\begin{align}
    \mathcal{I}_d^{\mathrm{full}}
    &= 2\mathrm{SNR}_0\,\beta_B,
    \label{eq:I-d-full}\\
    \boldsymbol{\mathcal{I}}_{k,k}^{\mathrm{full}}
    &=
    2\mathrm{SNR}_k
    \!\left[
    \!\left(1-\frac{|r_k|^2}{\|\mathbf{r}_{\mathrm{BD}}\|^2}\right)
    \mathbf{v}_k\mathbf{v}_k^\top
    +\beta_B\mathbf{1}_2\mathbf{1}_2^\top
    \right]\!,
    \label{eq:I-kk-full}\\
    \boldsymbol{\mathcal{I}}_{k,\ell}^{\mathrm{full}}
    &=
    -2\mathrm{SNR}_k
    \frac{|r_\ell|^2}
    {\|\mathbf{r}_{\mathrm{BD}}\|^2}
    \,\mathbf{v}_k\mathbf{v}_\ell^\top,
    \quad k\neq\ell.
    \label{eq:I-kl-full}
\end{align}
\end{subequations}
The off-diagonal expression satisfies
$\boldsymbol{\mathcal{I}}_{\ell,k}^{\mathrm{full}}
=(\boldsymbol{\mathcal{I}}_{k,\ell}^{\mathrm{full}})^\top$, so the full EFIM
remains symmetric.
Carrier-phase information is removed by unknown BD phases; amplitude-gradient
information is partially suppressed by the common gain/BD-coefficient ambiguity;
and the wideband delay term $\beta_B$ survives unchanged.
The proof is given in Appendix~\ref{app:efim-proofs}.
\end{corollary}

\subsection{Range-to-Position Mapping}
\label{sec:range-position-mapping}

The range-domain EFIMs above are bounds for the path-length vector
$\boldsymbol{\eta}_r$. To obtain a position-domain bound, we use the standard
FIM change of variables~\cite[Sec.~3.8]{kay1993fundamentalsStatistical}. Let
$\boldsymbol{\theta}\triangleq
[\mathbf{p}_{\mathrm{UE}}^\top,\mathbf{p}^\top]^\top$ denote the UE and
scatterer position vector. The deterministic map from positions to path lengths
is
\begin{equation}
\label{eq:range-position-map}
\begin{aligned}
    \boldsymbol{\eta}_r
    =
    \boldsymbol{\psi}(\boldsymbol{\theta})
    =
    \big[
    &\|\mathbf{p}-\mathbf{p}_{\mathrm{UE}}\|,\,
    \|\mathbf{p}^{(1)}-\mathbf{p}_{\mathrm{UE}}\|,\,
    \|\mathbf{p}^{(1)}-\mathbf{p}\|,\ldots,\\
    &\|\mathbf{p}^{(K)}-\mathbf{p}_{\mathrm{UE}}\|,\,
    \|\mathbf{p}^{(K)}-\mathbf{p}\|
    \big]^\top .
\end{aligned}
\end{equation}
The Jacobian $\boldsymbol{\mathcal{J}}(\boldsymbol{\eta}_r;
\mathbf{p}_{\mathrm{UE}},\mathbf{p})\triangleq
\partial\boldsymbol{\psi}(\boldsymbol{\theta})/
\partial\boldsymbol{\theta}^\top$ is
\begin{equation}
    \label{eq:jacobian-dist-positions}
    \begin{aligned}
    \boldsymbol{\mathcal{J}}(\boldsymbol{\eta}_r;
    \mathbf{p}_{\mathrm{UE}},\mathbf{p})
    &=
    \begin{bmatrix}
        \boldsymbol{\mathcal{J}}_{\mathrm{UE}} &
        \boldsymbol{\mathcal{J}}_{\mathrm{s}}
    \end{bmatrix} \\
    \end{aligned}
\end{equation}
where
$\boldsymbol{\mathcal{J}}_{\mathrm{UE}}
=\partial\boldsymbol{\eta}_r/\partial\mathbf{p}_{\mathrm{UE}}^\top$,
$\boldsymbol{\mathcal{J}}_{\mathrm{s}}
=\partial\boldsymbol{\eta}_r/\partial\mathbf{p}^\top$, and their non-zero rows are given by
$\mathbf{u}^\top(\mathbf{y},\mathbf{x})=(\mathbf{x}-\mathbf{y})^\top/\|\mathbf{x}-\mathbf{y}\|$.
For any calibration regime, let
$\boldsymbol{\mathcal{I}}(\boldsymbol{\eta}_r)$ denote the corresponding
range-domain EFIM after eliminating the appropriate nuisance parameters. Since
$\boldsymbol{\eta}_r=\boldsymbol{\psi}(\boldsymbol{\theta})$, the joint
UE/scatterer position EFIM is
\begin{equation}
\label{eq:position-efim}
    \boldsymbol{\mathcal{I}}(\boldsymbol{\theta})
    =
    \boldsymbol{\mathcal{J}}^\top(\boldsymbol{\eta}_r;
    \mathbf{p}_{\mathrm{UE}},\mathbf{p})
    \boldsymbol{\mathcal{I}}(\boldsymbol{\eta}_r)
    \boldsymbol{\mathcal{J}}(\boldsymbol{\eta}_r;
    \mathbf{p}_{\mathrm{UE}},\mathbf{p}) .
\end{equation}
When~\eqref{eq:position-efim} is nonsingular, the joint position CRB is
$\boldsymbol{\mathcal{I}}(\boldsymbol{\theta})^{-1}$.

For a UE-only bound, eliminating the scatterer coordinates by the Schur
complement gives the UE-position EFIM
\begin{equation}
\label{eq:ue-efim-scatterer-nuisance}
    \begin{aligned}
    \boldsymbol{\mathcal{I}}(\mathbf{p}_{\mathrm{UE}})
    &=
    \boldsymbol{\mathcal{J}}_{\mathrm{UE}}^\top
    \boldsymbol{\mathcal{I}}(\boldsymbol{\eta}_r)
    \boldsymbol{\mathcal{J}}_{\mathrm{UE}} \\
    &\!\!\!\!\!-
    \boldsymbol{\mathcal{J}}_{\mathrm{UE}}^\top
    \boldsymbol{\mathcal{I}}(\boldsymbol{\eta}_r)
    \boldsymbol{\mathcal{J}}_{\mathrm{s}}
    \left(
    \boldsymbol{\mathcal{J}}_{\mathrm{s}}^\top
    \boldsymbol{\mathcal{I}}(\boldsymbol{\eta}_r)
    \boldsymbol{\mathcal{J}}_{\mathrm{s}}
    \right)^{-1}
    \boldsymbol{\mathcal{J}}_{\mathrm{s}}^\top
    \boldsymbol{\mathcal{I}}(\boldsymbol{\eta}_r)
    \boldsymbol{\mathcal{J}}_{\mathrm{UE}} .
    \end{aligned}
\end{equation}
If the scatterer block is singular, the inverse
in~\eqref{eq:ue-efim-scatterer-nuisance} is interpreted as a generalized Schur
complement with the Moore--Penrose pseudoinverse. If the corresponding
range-space condition is not satisfied, the reduced EFIM is singular, indicating
that the UE position is not locally identifiable under that nuisance treatment.

The scatterer is eliminated by this Schur complement in every calibration
regime. The effective gain $g$ is treated as an independent complex coefficient;
although it physically absorbs the scatterer-to-BS segment
$d_{\mathrm{SB}}=\|\mathbf{p}_{\mathrm{BS}}-\mathbf{p}\|$ together with the
reflection, RF-chain, and synchronization factors, this position dependence is
deliberately excluded from the reduced model, and the EFIM differentiates only
the range coefficients $\rho_0,\rho_k$, not $g$. Calibration therefore means that
the complex coefficient is known, not that $\mathbf{p}$ is; the scatterer remains
a nuisance throughout, and the calibrated bound is conservative in that it does
not exploit any residual information about $\mathbf{p}$ that the physical gain
may carry.

Let $\mathbf{C}_{\mathrm{UE}}$ denote the CRB matrix associated with
$\mathbf{p}_{\mathrm{UE}}$. If the UE and scatterer positions are estimated
jointly, $\mathbf{C}_{\mathrm{UE}}$ is the
$\mathbf{p}_{\mathrm{UE}}$-block of
$\boldsymbol{\mathcal{I}}^{-1}(\boldsymbol{\theta})$. If the scatterer
position is treated as nuisance, then
$\mathbf{C}_{\mathrm{UE}}=\boldsymbol{\mathcal{I}}^{-1}
(\mathbf{p}_{\mathrm{UE}})$ with
$\boldsymbol{\mathcal{I}}(\mathbf{p}_{\mathrm{UE}})$ given
by~\eqref{eq:ue-efim-scatterer-nuisance}.
The UE position error bound (PEB) is then formally defined as
\begin{equation}
\label{eq:peb}
    \mathrm{PEB}
    =
    \sqrt{\operatorname{tr}(\mathbf{C}_{\mathrm{UE}})} .
\end{equation}
The PEB is a scalar lower bound on the root-mean-square Euclidean position
error of any unbiased UE-position estimator; equivalently, the trace in
\eqref{eq:peb} sums the CRB contributions along the UE-position coordinates.

\subsection{Discussion}

We next interpret the modeling assumptions, nuisance-parameter losses,
identifiability conditions, and RIS/subpanel calibration regimes implied by the
closed-form EFIMs.

\subsubsection{Modeling Assumptions}

The central modeling assumption is a single dominant scatterer, which underlies
the common gain nuisance. The results are nonetheless more general than this
global presentation suggests. The common-scatterer assumption can hold locally,
with subsets of BDs sharing one dominant scatterer-to-BS interaction while
different subsets are tied to different scatterers. The resulting EFIM keeps the
same Schur-complement structure, with coupling restricted to BDs sharing the
same gain nuisance.

\begin{remark}[Grouped path gains]
\label{rem:grouped-path-gains}
When the BD-assisted gains are not described by a single common coefficient, the
same Schur-complement structure applies group-wise. As an example, suppose the
BDs are partitioned into two disjoint groups $\mathcal{G}_1$ and $\mathcal{G}_2$
with
\begin{equation}
    z_k=\alpha_i r_k+w_k,\qquad k\in\mathcal{G}_i,\quad i\in\{1,2\},
\end{equation}
where $\alpha_i$ is an unknown complex gain shared only within group
$\mathcal{G}_i$. The direct NLOS component need not be absent. When observable,
it is assigned to whichever group shares its gain, so that its range parameter
$d$ joins that subgroup block and couples only with the BD ranges there; when
absent, $d$ and its EFIM terms are simply dropped. Ordering the range parameters
by group, the EFIM is block diagonal,
\begin{equation}
    \boldsymbol{\mathcal{I}}^{(\alpha_1,\alpha_2)}
    =
    \begin{bmatrix}
    \boldsymbol{\mathcal{I}}_{\mathcal{G}_1}^{(\alpha_1)} & \mathbf{0}\\
    \mathbf{0} & \boldsymbol{\mathcal{I}}_{\mathcal{G}_2}^{(\alpha_2)}
    \end{bmatrix},
\end{equation}
each block being reduced exactly as in the single-group problem, with the
BD-only part following Proposition~\ref{prop:unknown-gain-bd-coefficient-efim}
when the group contains no calibrated reference, and
Corollary~\ref{cor:unknown-common-gain} when it also contains the direct NLOS
component with a known relative response. A block thus has dimension
$2|\mathcal{G}_i|$ when it contains only BD-assisted ranges, or
$2|\mathcal{G}_i|+1$ when it also contains the direct range $d$. Coupling
therefore arises only among paths that share the same unknown group gain, while
cross-group blocks vanish.

In the most fragmented case, when every group is a singleton, each
BD-assisted path $k=1,\ldots,K$ carries its own independent gain $\alpha_k$,
whose coefficient-gradient information is absorbed by that gain. Eliminating the
$\alpha_k$ by the Schur complement removes every direct--BD and inter-BD
gain-induced coupling, leaving only the per-path waveform-delay contribution
(plus any externally calibrated coefficients or phases). The shared-gain results
are thus conditional on a common or group-wise dominant scatterer, and this fully
independent case is their pessimistic worst-case limit, retaining only per-path
delay information when fully uncalibrated.
\end{remark}

The direct NLOS component plays a separate role, helping separate the common
scatterer-to-BS gain $g$ from the relative BD reflection/modulation coefficient
$\gamma$.

\begin{remark}[Absence of the direct NLOS component]
\label{rem:absent-direct-nlos}
The derivation assumes a resolvable direct NLOS component when $g$ and
$\gamma$ are interpreted as separate nuisance parameters. If this component is
absent, the BD-assisted observations identify only the product
$\alpha=g\gamma$. The same EFIM structure applies after reparameterizing the
BD-assisted paths by the single common complex gain $\alpha$ and removing the
direct-path range parameter $d$ and its EFIM terms.

In the position-domain EFIM, this removal deletes the relative UE--scatterer
range constraint associated with
$d=\|\mathbf{p}-\mathbf{p}_{\mathrm{UE}}\|$. Consequently, the loss is often
most visible in the scatterer-position block and in the UE--scatterer coupling.
The UE may still be constrained by UE--BD ranges, whereas the scatterer loses
one direct constraint tying it to the UE.
\end{remark}

The measurement model requires the direct NLOS and BD-assisted components to be
resolvable and correctly associated in delay and Doppler, as discussed next.

\begin{remark}[Component association and separability]
\label{rem:component-association}
The known BD signatures determine the Doppler-bin association, whereas delay-bin
association may additionally use bandwidth-resolved peaks and candidate
geometry.
General data association in dynamic multipath is outside the present scope.
Weak unresolved multipath, off-grid leakage, and known delay-bin phase rotations
are absorbed into the effective gains and disturbances. The resulting EFIMs are
therefore genie-aided with respect to component separation; imperfect
association, finite-aperture leakage, or nonorthogonal signatures can only
reduce the available information.
\end{remark}

The calibrated EFIM retains the full carrier-phase contribution and therefore
contains the largest amount of localization information. The resulting CRB is
optimistic, as elaborated next.

\begin{remark}[Continuous carrier-phase bound]
\label{rem:continuous-phase-bound}
The carrier-phase terms in the EFIMs are continuous-phase bounds. A
carrier-phase observation is ambiguous modulo $2\pi$ and therefore does not
determine an absolute delay unless the integer number of carrier cycles is
resolved. Mixed-integer CRB analyses such as~\cite{wymeersch2023carrierPhaseBounds} show
that when the delay-domain uncertainty is large relative to $\lambda$, this
integer ambiguity cannot be resolved and the attainable bound approaches the
delay-only limit. Therefore, the carrier-phase gains predicted by the
calibrated EFIMs should be interpreted as optimistic bounds, attainable only
when cycle ambiguities are resolved by sufficient delay accuracy, prior
information, mobility, or other side information. In regimes where $g$,
$\gamma$, or $\boldsymbol{\phi}$ are treated as calibrated, the corresponding
complex phases must also be known with a phase reference accurate enough to make
the carrier phase meaningful.
\end{remark}

\subsubsection{Nuisance Parameters}

In some controlled deployments, $g$ can be regarded as calibrated or learned
from auxiliary measurements. In practical passive-BD operation, however, this
calibration is difficult because $g$ absorbs the unknown scatterer-to-BS
response, propagation loss, and receiver-chain effects. Therefore, the
unknown-$g$ and fully unknown regimes are the most relevant baselines.
When the direct NLOS component is observable and $g$ is unknown, the effect of
gain elimination depends on whether the BD-assisted paths provide a calibrated
relative reference.

\begin{remark}[Calibration-loss mechanisms and direct--BD coupling]
\label{rem:direct-bd-gain-coupling}
\label{rem:bw-floor}
\label{rem:cascade}
\label{rem:carrier-phase}
The EFIM degrades through phase erasure and shared-nuisance coupling.

\emph{Phase erasure}: unknown $\boldsymbol{\phi}$ removes the
$(2\pi/\lambda)^2$ carrier-phase term from every BD-assisted block
(Proposition~\ref{prop:unknown-phase-efim}; compare~\eqref{eq:I-k-cal}
and~\eqref{eq:I-k-phi}). Marginalizing uniformly random phases leads to the same
qualitative loss, although the resulting noncoherent likelihood is not identical
to the deterministic-nuisance EFIM at finite SNR. Both cases differ from the
cycle ambiguity of Remark~\ref{rem:continuous-phase-bound}, where continuous
phase is available but its integer cycle count is unresolved.

\emph{Shared-nuisance coupling}: eliminating a nuisance shared across several
paths introduces off-diagonal EFIM blocks and suppresses their common derivative
directions. In particular, when $\gamma$ and $\boldsymbol{\phi}$ are calibrated
but $g$ is unknown,
Corollary~\ref{cor:unknown-common-gain} shows that the reduced range EFIM retains
the direct--BD blocks $\boldsymbol{\mathcal{I}}_{d,k}^{(g)}$
in~\eqref{eq:I-dk-g-only}. These arise because the direct range $d$ and the
BD-assisted ranges both carry cross-information with the real and imaginary parts
of $g$ in the unreduced partition~\eqref{eq:EFIM-Schur}, so eliminating $g$ is a
joint Schur reduction over the direct and BD-assisted observations: part of the
direct-path narrowband information survives through relative direct/BD
measurements.

This coupling is what protects the direct path. When it is absent---either
because $\gamma$ is also unknown, so that the enlarged nuisance block of
Proposition~\ref{prop:unknown-gain-bd-coefficient-efim} removes the direct--BD
blocks, or because no calibrated relative BD reference is available---the direct
NLOS path retains only the bandwidth-dependent term $\beta_B$,
see~\eqref{eq:I-d-gain} and~\eqref{eq:I-d-full}. A narrowband waveform
($\beta_B\to0$) then leaves no absolute direct-path range information, the
worst-case gain-uncertain regime.
\end{remark}

When carrier-phase information is removed by nuisance phases or limited by
cycle ambiguity, the range EFIM still contains two non-phase contributions:
the wideband-delay term and the amplitude-gradient term.
The bandwidth term is controlled by $\beta_B$, whereas the
amplitude-gradient term is governed by $\mathbf{v}_k\mathbf{v}_k^\top$ and
therefore depends on the per-BD SNR and on the UE--BD/scatterer--BD distance
derivatives. Unlike carrier phase, these terms do not require resolving an
integer number of carrier cycles, but they are typically weaker and more
geometry dependent.

\subsubsection{Identifiability Consequences}
\label{sec:identifiability}

Joint single-snapshot local identifiability of
$\boldsymbol{\theta}=[\mathbf{p}_{\mathrm{UE}}^\top,\mathbf{p}^\top]^\top$
requires $\mathrm{rank}\{\boldsymbol{\mathcal{I}}(\boldsymbol{\theta})\}=\dim(\boldsymbol{\theta})$,
where $\boldsymbol{\mathcal{I}}(\boldsymbol{\theta})$ is the position EFIM
from~\eqref{eq:position-efim}. In two dimensions $\dim(\boldsymbol{\theta})=4$;
in three dimensions it is $6$. This rank condition requires both enough
independent range directions after nuisance elimination and a geometry Jacobian
that maps those directions into distinct position updates.

The first requirement concerns the range-domain EFIM. In the fully unknown
regime, a single BD is already insufficient before the range information is
mapped to position.

\begin{remark}[$K=1$ singularity]
\label{rem:K1}
With a single BD, $\|\mathbf{r}_{\mathrm{BD}}\|^2=|r_1|^2$, so the
attenuation factor $1-|r_1|^2/\|\mathbf{r}_{\mathrm{BD}}\|^2=0$.
In the fully unknown regime~\eqref{eq:I-kk-full}, the $2\times2$ BD block
reduces to the rank-one matrix $\beta_B\mathbf{1}_2\mathbf{1}_2^\top$, and the
two segment lengths $d_{\mathrm{UE}}^{(1)}$ and $d^{(1)}$ cannot be separated.
Thus a single BD is insufficient in the presence of gain and phase nuisance,
even before the position-domain geometric rank conditions are considered.
\end{remark}

Even with several BDs, the two segment lengths of each assisted path must be
distinguishable within its range-domain block.

\begin{remark}[BD-block conditioning]
\label{rem:bd-block-conditioning}
Already in the calibrated case, each BD-assisted block
in~\eqref{eq:I-k-cal} is a sum of two rank-one directions,
$\mathbf{v}_k\mathbf{v}_k^\top$ and
$\mathbf{1}_2\mathbf{1}_2^\top$. The sum of these two rank-one matrices has
full column rank when the directions $\mathbf{v}_k$ and $\mathbf{1}_2$ are
linearly independent. Since
$\mathbf{v}_k=[1/d_{\mathrm{UE}}^{(k)},1/d^{(k)}]^\top$, exact rank deficiency
occurs when $d_{\mathrm{UE}}^{(k)}=d^{(k)}$, while near equality leads to poor
conditioning. Hence, the two segment lengths $d_{\mathrm{UE}}^{(k)}$ and
$d^{(k)}$ are locally distinguishable only through the non-collinearity of
$\mathbf{v}_k$ and $\mathbf{1}_2$.
\end{remark}

The second requirement concerns the range-to-position geometry. Under favorable
range-domain conditioning, at least two BDs in two dimensions and three BDs in
three dimensions are necessary for joint single-snapshot local identifiability
of the UE and scatterer positions. These counts are not universal thresholds for
UE-only localization with prior information or multiple snapshots, nor are they
sufficient, since the Jacobian in~\eqref{eq:jacobian-dist-positions} must have full
column rank. Small BD angular spread makes it ill-conditioned even when full rank
holds~\cite{sadeghi2021optimalGeometry}.

For UE-only localization, the unknown scatterer is then eliminated through
\eqref{eq:ue-efim-scatterer-nuisance}; its conditioning determines how much
position information is lost to this geometric nuisance.

\begin{remark}[Scatterer nuisance conditioning]
\label{rem:scatterer-nuisance-rank}
If
$\boldsymbol{\mathcal{J}}_{\mathrm{s}}^\top
\boldsymbol{\mathcal{I}}(\boldsymbol{\eta}_r)
\boldsymbol{\mathcal{J}}_{\mathrm{s}}$ is singular, some scatterer-displacement
directions are unidentifiable from the available ranges, and the inverse
in~\eqref{eq:ue-efim-scatterer-nuisance} becomes a Moore--Penrose pseudoinverse
that eliminates only the identifiable nuisance subspace. Exact rank deficiency
occurs in the collinear geometry, with the UE, scatterer, and BDs on one line:
the rows $\mathbf{u}(\mathbf{p}_{\mathrm{UE}},\mathbf{p})^\top$ and
$\mathbf{u}(\mathbf{p}^{(k)},\mathbf{p})^\top$ are then all parallel and
$\boldsymbol{\mathcal{J}}_{\mathrm{s}}$ loses full column rank. A near-degenerate
case arises when the UE and BDs are far from the scatterer relative to their
spread, so the direction vectors become nearly parallel and the scatterer is
weakly constrained in cross-range.
\end{remark}

The rank conditions above are local, and full column rank of
$\boldsymbol{\mathcal{J}}$ and a nonsingular EFIM guarantee only that the
position parameters are locally identifiable in a neighborhood of
$\boldsymbol{\theta}$. Geometric symmetries of the deployment can introduce
discrete global ambiguities even when every local condition is satisfied.

\begin{remark}[Global ambiguity]
\label{rem:global-ambiguity}
Full column rank of $\boldsymbol{\mathcal{J}}$ does not preclude discrete
global ambiguities: two distinct coordinate pairs
$(\mathbf{p}_{\mathrm{UE}},\mathbf{p})$ can produce the same distance vector
$\boldsymbol{\eta}_r$, yielding the same likelihood and rendering the global
solution non-unique. For example, when the BDs are collinear, jointly reflecting
the UE and scatterer across the BD line preserves all ranges in
$\boldsymbol{\eta}_r$~\cite{torrieri1984statisticalPassive}.
Resolving such ambiguities requires either geometric diversity that breaks the
symmetry (e.g., non-collinear BD placement) or prior information on the
deployment.
\end{remark}

\subsubsection{RIS Calibration Interpretation}

Section~\ref{sec:bd-ris-equivalence} emphasized that the BD-style vector model
applies to a RIS only when element-wise or subpanel-wise reflected components
are separately observable. Under this separability assumption, the EFIM results
above can be read as a calibration dictionary for separable
RIS/subpanel-assisted NLOS positioning, not for a generic coherent-aperture RIS
measurement. The relevant distinction is not whether the passive device is a BD
or a RIS subpanel, but which complex coefficients and residual phases are known
to the estimator.

\begin{remark}[RIS calibration interpretation]
\label{rem:ris-calibration}
For a separable RIS, the calibration regimes map onto the EFIM results---and
onto practical deployments---as follows:
\begin{enumerate}[wide=0pt, label=(\roman*)]
    \item \emph{Fully calibrated} (gain, relative response, and element/subpanel
    phases known; Proposition~\ref{prop:calibrated-efim}): engineered anchors
    characterized in advance, such as a controller-driven RIS subpanel or a
    laboratory-characterized BD.
    \item \emph{Unknown phase only} (only residual phases unknown;
    Proposition~\ref{prop:unknown-phase-efim}): a synchronized panel or
    co-clocked BDs with stable amplitudes but residual phases that drift through
    phase quantization, synchronization offsets, or temperature, so relative
    gains stay trustworthy while absolute phase does not.
    \item \emph{Unknown common gain} (relative response and phases calibrated,
    common gain unknown; Corollary~\ref{cor:unknown-common-gain}).
    \item \emph{Unknown panel coefficient} (gain calibrated, a single common
    reflection coefficient unknown, relative phases known;
    Corollary~\ref{cor:individual-g-gamma}).
    \item \emph{Unknown gain and coefficient} (common gain and coefficient
    unknown, relative phases known;
    Proposition~\ref{prop:unknown-gain-bd-coefficient-efim}): low-cost hardware
    whose reflection efficiency and receive-chain gain are uncharacterized
    (e.g., unknown antenna-mode scattering), while residual phases can still be
    referenced.
    \item \emph{Fully uncalibrated} (unknown per-path phases atop unknown gains;
    Corollary~\ref{cor:fully-unknown-efim}): off-the-shelf ambient BDs deployed
    opportunistically with no calibration of any reflection, phase, or channel
    term---the most representative Ambient IoT case and the most pessimistic
    baseline.
\end{enumerate}
The gap between the calibrated and fully uncalibrated cases therefore quantifies the
value of phase and gain calibration, while the intermediate results distinguish
the contribution of relative-phase knowledge from that of common-gain knowledge.
\end{remark}

Beyond the information-level degradation captured by the calibration hierarchy,
a single planar RIS can also provide limited geometric diversity because its
element positions $\mathbf{p}^{(k)}$ are confined to a plane (or line in 2D).
When the aperture subtends only a small angle at the scatterer,
$\boldsymbol{\mathcal{J}}$ becomes ill-conditioned in the manner identified in
Remark~\ref{rem:bd-block-conditioning}. In an extreme planar-collinear geometry,
the Jacobian loses full column rank entirely, reproducing the degeneracy of
Remark~\ref{rem:scatterer-nuisance-rank} and potentially introducing the global
mirror-image ambiguity of Remark~\ref{rem:global-ambiguity}. This is not a
limitation of planarity alone; its severity depends on aperture size,
orientation, propagation geometry, and whether multiple surfaces are available.
Calibration cannot repair a rank-deficient deployment, so these geometric
factors must be considered in RIS placement.

\section{Numerical Results}
\label{sec:numerical-results}

\begin{table}[t]
\caption{Baseline parameters for numerical evaluation}
\label{tab:simulation-scenario}
\centering
\small
\setlength{\tabcolsep}{3pt}
\renewcommand{\arraystretch}{1.15}
\begin{tabularx}{\columnwidth}{@{}L{0.34\columnwidth}X@{}}
\toprule
\textbf{Item} & \textbf{Baseline choice} \\
\midrule
Deployment geometry &
  $4$~m-wide corridor-to-room NLOS geometry;
  \begin{tabular}[t]{@{}l@{}}
  $\mathbf{p}_{\mathrm{UE}}=(0,0)$~m,
  $\mathbf{p}=(-3.6,1.4)$~m,\\
  $\mathbf{p}_{\mathrm{BS}}=(-2.8,4.0)$~m
  \end{tabular} \\
BD placement &
  \begin{tabular}[t]{@{}l@{}}
  $\mathbf{p}^{(1)}=(-0.4,2.0)$~m,
  $\mathbf{p}^{(2)}=(0.6,2.0)$~m,\\
  $\mathbf{p}^{(3)}=(0.8,-2.0)$~m,
  $\mathbf{p}^{(4)}=(-1.2,-2.0)$~m
  \end{tabular} \\
Carrier frequency &
  $f_c=3.5$~GHz \\
SRS numerology &
  \begin{tabular}[t]{@{}l@{}}
  $\Delta_f=30$~kHz,\\
  $K_{\mathrm{TC}}=4$,\\
  $m_{\mathrm{SRS},b}=120$~PRBs,\\
  $M_{\mathrm{SRS}}=360$ active tones,\\
  $B_{\mathrm{rms}}=12.47$~MHz
  \end{tabular} \\
Direct-NLOS SNR &
  $20$~dB per resolved bin \\
Relative BD coefficient &
  $|\gamma|=1$~m (hardware-identical BDs) \\
\bottomrule
\end{tabularx}
\renewcommand{\arraystretch}{1}
\end{table}

This section numerically evaluates the EFIM cases derived above to illustrate
the information-loss mechanisms, identifiability conditions, and geometric
effects analyzed in Section~\ref{sec:crb-results}.

\subsection{Simulation Scenario}
\label{sec:simulation-scenario}

The numerical studies use the corridor-to-room NLOS deployment and waveform
parameters summarized in Table~\ref{tab:simulation-scenario} and illustrated in
Fig.~\ref{fig:system-model}. The default opposite-wall BD placement provides
non-collinear bearings, placing the scenario above the geometric degeneracy of
Remark~\ref{rem:scatterer-nuisance-rank} while keeping the deployment passive.
Only the BD placement is varied in the $K$-sweep and geometric-sensitivity
figures, according to the placement rules stated with the corresponding
figures. These scenarios are designed to expose the theoretically predicted
nuisance-parameter losses and geometric effects, rather than to minimize PEB.
For the PEB, the scatterer is eliminated as a nuisance in all regimes via the
Schur complement~\eqref{eq:ue-efim-scatterer-nuisance}.

The SRS parameters in Table~\ref{tab:simulation-scenario} use a mid-band
carrier and a mid-range uplink sounding allocation
($K_{\mathrm{TC}}=4$, $m_{\mathrm{SRS},b}=120$~PRBs), yielding a finite RMS
bandwidth $B_{\mathrm{rms}}=12.47$~MHz, or equivalently
$2\pi B_{\mathrm{rms}}/c=0.261$~rad/m, that reveals the delay-dependent
bandwidth floor without making it dominate the carrier-phase and
amplitude-gradient effects.

For all curves, $\mathrm{SNR}_i\triangleq|\mu_i|^2/\sigma_w^2$ denotes the
post-processing SNR of the selected delay--Doppler bin. Unless otherwise stated,
the direct-NLOS bin has $\mathrm{SNR}_0=20$~dB. This definition includes the
coherent slow-time FFT gain; hence, the FFT length affects only the raw-sample
SNR required to obtain $\mathrm{SNR}_i$ and does not otherwise enter the EFIM.

The BD-bin SNRs are scaled according to their relative geometric path losses:
\begin{equation}
    \mathrm{SNR}_k
    =
    \mathrm{SNR}_0|\gamma|^2
    \frac{d^2}
    {\bigl(d_{\mathrm{UE}}^{(k)}d^{(k)}\bigr)^2}.
\end{equation}
Here we use $|\gamma|=1$~m, which makes the ratio
dimensionless under the adopted path normalization. Common gain factors are
absorbed into $\mathrm{SNR}_0$, while geometry enters the EFIM through both the
BD-bin SNRs and the distance derivatives in~\eqref{eq:I-d-cal}
and~\eqref{eq:I-k-cal}.

The absolute PEB values reported below are tied to the scenario in
Table~\ref{tab:simulation-scenario}; path lengths, SNR scaling, bandwidth,
carrier frequency, and BD placement must be recomputed for each deployment.
The qualitative conclusions are structural: phase uncertainty removes BD
carrier-phase information, shared-response uncertainty controls cross-BD
coupling, and geometric diversity determines how range information maps into the
UE-position EFIM.

\subsection{Ranging Information and PEB Results}
\label{sec:numerical-evaluation}

\begin{figure*}[!t]
\centering
\subfloat[]{%
    \includegraphics[width=0.33\textwidth]{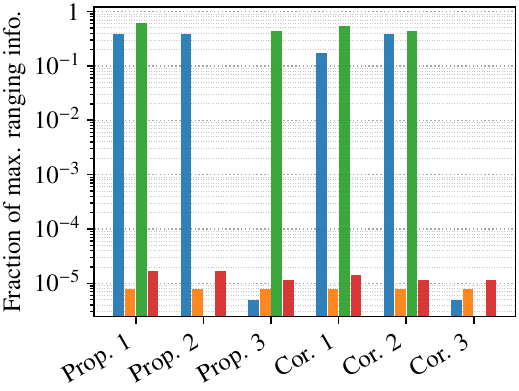}
    \label{fig:range-info-breakdown}}\hspace{0.015\textwidth}
\subfloat[]{%
    \includegraphics[width=0.30\textwidth]{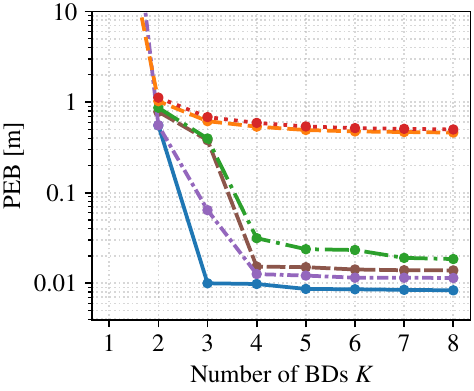}
    \label{fig:peb-vs-k}}\hspace{0.015\textwidth}
\subfloat[]{%
    \includegraphics[width=0.30\textwidth]{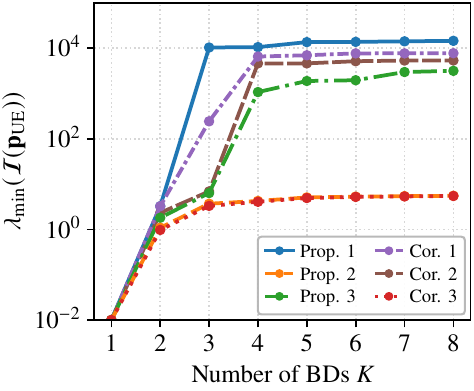}
    \label{fig:lmin-vs-k}}
\caption{(a)~Range-domain information allocation across the EFIM cases,
normalized by the total information in Proposition~\ref{prop:calibrated-efim}.
Bars denote: \figswatch{1f77b4}~direct NLOS;
\figswatch{ff7f0e}~BD wideband-delay;
\figswatch{2ca02c}~BD carrier-phase; and
\figswatch{d62728}~BD amplitude-gradient.
(b)~PEB versus number of BDs $K$ at 20~dB direct-NLOS SNR, using cumulative BD
placement on both walls; nonfinite values are omitted.
(c)~Minimum eigenvalue $\lambda_{\min}(\boldsymbol{\mathcal{I}}(\mathbf{p}_{\mathrm{UE}}))$
of the reduced UE-position EFIM versus $K$ at 20~dB direct-NLOS SNR, after
eliminating the scatterer as nuisance; values plotted at $10^{-2}$ represent
nonpositive or below-floor eigenvalues. The line legend in (c) applies to
panels~(b) and~(c).}
\label{fig:peb-conditioning}
\end{figure*}

Fig.~\ref{fig:range-info-breakdown} isolates the range-domain information
allocation for the baseline scenario at $20$~dB direct-NLOS SNR. The direct
block and the summed traces of the BD wideband-delay, carrier-phase, and
amplitude-gradient blocks are normalized by the total calibrated information.
This trace decomposition is a range-domain diagnostic rather than a localization
bound. It visualizes the information-loss mechanisms derived above. Phase uncertainty
removes the BD carrier-phase contribution, whereas joint gain and relative-
coefficient uncertainty leaves only the direct-path bandwidth term and
attenuates the BD narrowband contributions. The intermediate regimes retain
different subsets of these components according to their calibrated parameters.

Fig.~\ref{fig:peb-vs-k} maps these range-domain losses to the position domain.
BDs are placed cumulatively starting from the BD1 position in
Fig.~\ref{fig:system-model}, alternating between the top and bottom corridor
walls (y\,=\,$\pm2$\,m) with $1$\,m x-increments; the direct-NLOS SNR is fixed
at $20$~dB and BD SNRs are scaled geometrically with path length. The calibrated
and phase-referenced regimes improve rapidly with $K$, whereas phase-uncertain
regimes plateau near $0.5$~m. At $K=4$, for example, the calibrated PEB is
$0.98$~cm, compared with $54$~cm when the BD phases are unknown. Additional
BDs therefore provide their largest benefit when their carrier phases are
available; the centimetre-level values remain optimistic continuous-phase
bounds subject to the ambiguity conditions of
Remark~\ref{rem:continuous-phase-bound}.

Fig.~\ref{fig:lmin-vs-k} uses the same BD placement and SNR to examine
UE-position conditioning through the smallest eigenvalue of the reduced EFIM
$\boldsymbol{\mathcal{I}}(\mathbf{p}_{\mathrm{UE}})$ after the scatterer is
eliminated. This metric follows the same range-to-position mapping and nuisance
elimination used for the PEB, so it measures the weakest local UE-position
information direction rather than an abstract null direction in the unconstrained
range coordinates. The curves show that the unknown-$\gamma$ and joint
unknown-$\{g,\gamma\}$ regimes remain locally identifiable once enough
geometrically distinct BDs are present, although their worst-direction
information remains below the calibrated and phase-referenced regimes. The
phase-uncertain regimes have the smallest eigenvalues because the BD
carrier-phase term is removed before the position-domain mapping.

\begin{figure}[!t]
\centering
\includegraphics[width=0.7\columnwidth]{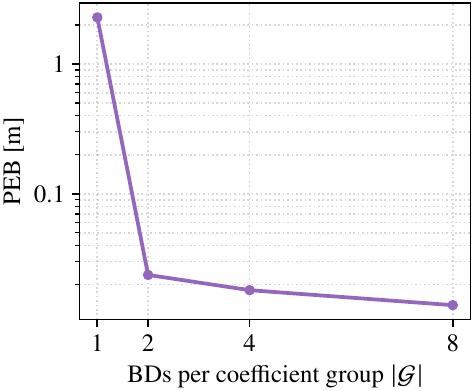}
\caption{PEB versus the group size $|\mathcal{G}|$, i.e., the number of BDs
sharing each independently unknown composite response coefficient, for $K=8$
partitioned into equal interleaved groups, with fixed geometry and per-bin SNRs.}
\label{fig:peb-vs-gamma-groups}
\end{figure}

Fig.~\ref{fig:peb-vs-gamma-groups} tests the shared composite-response
approximation of Remark~\ref{rem:grouped-path-gains} by partitioning the $K=8$
BDs into equal, spatially interleaved groups and varying the group size
$|\mathcal{G}|\in\{1,2,4,8\}$, the number of BDs sharing each independently
unknown group response. In the implementation this response is represented as a
group-specific relative coefficient $\gamma_g$ with known $g$, but this is
algebraically equivalent to a group-specific common scatterer-to-BS gain
$g_g$ with a fixed relative coefficient, since only the product enters the
BD-assisted mean. The geometry and per-bin SNRs are fixed, while
$\boldsymbol{\phi}$ remains known, so the change reflects only the loss of
cross-BD response sharing; all eight BDs contribute geometric information at
every $|\mathcal{G}|$, and only the shared-response structure changes. The PEB falls
from $2.29$~m when each BD has its
own coefficient ($|\mathcal{G}|=1$) to $1.39$~cm when all eight share one
($|\mathcal{G}|=8$), showing that the single-snapshot narrowband information
relies strongly on shared response structure.

\subsection{Geometric Sensitivity}
\label{sec:geometric-sensitivity}

\begin{figure*}[t]
\centering
\begin{tabular*}{\textwidth}{@{\extracolsep{\fill}}ccc@{}}
\subfloat[]{%
    \includegraphics[height=1.45in, width=0.25\textwidth]{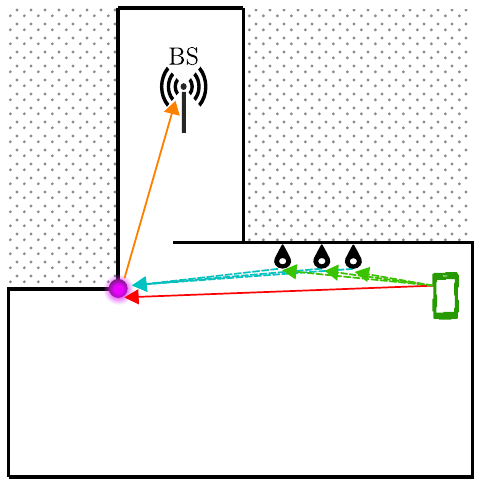}
    \label{fig:bad-geometry-scenario}} &
\subfloat[]{%
    \includegraphics[height=1.45in, width=0.31\textwidth]{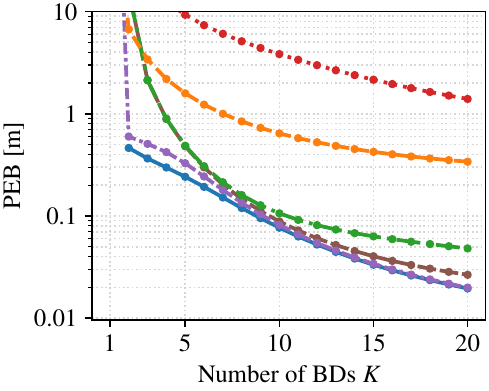}
    \label{fig:peb-bad-k}} &
\subfloat[]{%
    \includegraphics[height=1.45in, width=0.36\textwidth]{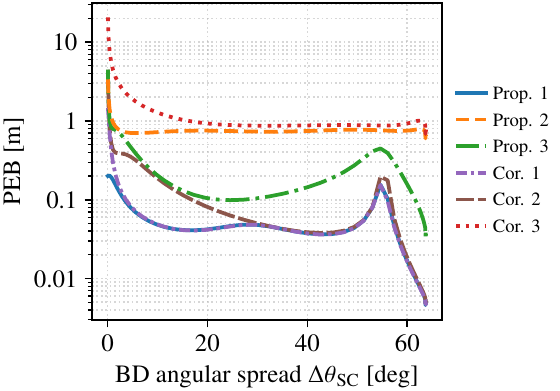}
    \label{fig:peb-angular}}
\end{tabular*}
\caption{(a)~Representative bad-geometry scenario: all BDs collinear on the
same corridor wall, providing poor angular diversity.
(b)~PEB versus number of BDs $K$ (up to $K\!=\!20$) for the collinear
$\lambda$-spaced placement on the top wall starting from BD1, at 20~dB direct-NLOS SNR.
(c)~PEB versus the resulting BD angular spread
$\Delta\theta_{\mathrm{SC}}$ as seen from the scatterer for $K\!=\!4$, with all
BD SNRs held equal to the direct-NLOS SNR to isolate geometry.
The line legend shown in (c) applies to both (b) and (c).}
\label{fig:geometric-sensitivity}
\end{figure*}

Fig.~\ref{fig:geometric-sensitivity} separates two geometric effects in the
position-domain bounds. First, Fig.~\ref{fig:bad-geometry-scenario} and
Fig.~\ref{fig:peb-bad-k} show what happens when additional BDs are added along
a nearly collinear deployment. Second, Fig.~\ref{fig:peb-angular} isolates
bearing diversity by sweeping the angular spread while holding the BD SNRs
fixed. Together, these experiments test the position-domain consequence of the
Jacobian conditioning discussed in
Remark~\ref{rem:scatterer-nuisance-rank}.

In the bad-geometry deployment of Fig.~\ref{fig:bad-geometry-scenario}, all BDs
are placed on the same corridor wall, so their directions as seen from the
scatterer are nearly collinear. Fig.~\ref{fig:peb-bad-k} evaluates this
configuration with wavelength-spaced BDs up to $K\!=\!20$, using the same
20~dB direct-NLOS SNR as before. This placement resembles a linear RIS with
elements spaced at $\lambda$ intervals. The PEB improves only slowly with
$K$, because adding more BDs along the same line mostly adds range equations
with nearly dependent geometric directions. As an illustrative contrast, for
Corollary~\ref{cor:individual-g-gamma} (unknown~$\gamma$), the two-wall
deployment used in Fig.~\ref{fig:peb-vs-k} reaches centimetre-level accuracy
by $K\!=\!4$ ($\approx\!1.5$~cm) when the residual phases are known, whereas
the collinear placement remains at $\approx\!2.7$~cm even at
$K\!=\!20$---so even twenty collinear BDs do not match four
well-separated ones.

Fig.~\ref{fig:peb-angular} then removes the confounding effect of path loss to
isolate bearing diversity. We place $K\!=\!4$ BDs at equal angular increments
$\delta_{\theta}$ on a UE-centred arc of radius equal to the BD1--UE distance,
starting from the BD1 position in Fig.~\ref{fig:system-model}. All BD SNRs are
held equal to the direct-NLOS SNR, so the sweep changes the BD bearing geometry
without changing the received BD powers; the horizontal axis reports the
resulting BD angular spread $\Delta\theta_{\mathrm{SC}}$ as seen from the
scatterer. The PEB degrades sharply as $\Delta\theta_{\mathrm{SC}}$ decreases,
with a steep cliff as the geometry approaches collinearity. The curve is not
strictly monotone in angular spread alone: local degradations occur when the
absolute BD bearings and ranges make the UE/scatterer Jacobian poorly
conditioned. Thus angular spread is a useful geometric summary, but the full
position-domain conditioning depends on the complete BD placement. These
results support the identifiability discussion: the number of resolvable BD
paths is necessary, but their geometric diversity controls whether the range
information maps into a well-conditioned position-domain EFIM.

\section{Conclusion}
\label{sec:conclusion}

This paper developed a CRB framework for BD-assisted NLOS localization under a
dominant scatterer-to-BS interaction shared by the direct and BD-assisted paths.
From a compact range-domain model, we derived closed-form EFIMs for calibrated,
partially calibrated, and fully uncalibrated operation; the same expressions
apply to separable RIS/subpanel-assisted links with individually observable
reflected components.

The results show how calibration and geometry set the useful ranging
information. Unknown residual phases remove carrier-phase ranging, unknown shared
gain or relative-coefficient nuisances induce Schur-complement losses and
couplings, and finite SRS bandwidth supplies the residual delay term. The
BD-assisted information also relies on shared response structure across devices,
degrading gracefully as that sharing is lost and collapsing only when every BD is
independently uncalibrated. In the
position domain, scatterer nuisance coupling and BD angular diversity further
limit this information, so adding passive paths does not help if their geometry
is poorly conditioned; joint single-snapshot localization further requires at
least two BDs in 2D (three in 3D) with sufficiently diverse bearings to the
scatterer, so geometric diversity, not BD count alone, governs the position
bound. The numerical examples confirm that centimetre-level
continuous-phase bounds require resolved carrier-cycle ambiguities and
phase-referenced calibration. Practical deployments should therefore prioritize
resolvable passive signatures, phase calibration, sufficient sounding bandwidth,
and diverse anchor bearings to the dominant scatterer.

\appendices

\section{OFDM/SRS Delay-Domain Measurement}
\label{appendix:measurement-extraction}

This appendix details the standard OFDM/SRS processing that leads to the
delay-domain observation in~\eqref{eq:delay-domain-channel}. The UE transmits a
known 5G NR SRS on a set of subcarriers and OFDM symbols. Let $n$ denote the
subcarrier index and $\ell$ the OFDM-symbol index. After cyclic-prefix removal
and the OFDM FFT, the received frequency-domain sample is
\begin{equation}
\label{eq:supp-ofdm-input-output}
    \widetilde{Y}[n,\ell]
    =
    H[n,\ell]S_{\mathrm{SRS}}[n,\ell]
    +\widetilde{W}[n,\ell],
\end{equation}
where $S_{\mathrm{SRS}}[n,\ell]$ is the known pilot,
$H[n,\ell]$ is the frequency-domain channel response, and
$\widetilde{W}[n,\ell]$ is receiver noise.

For the constant-amplitude SRS,
$|S_{\mathrm{SRS}}[n,\ell]|^2=\mathcal{P}_s$, pilot removal gives
\begin{equation}
\label{eq:supp-descrambled-signal}
\begin{aligned}
    Y[n,\ell]
    &=
    \widetilde{Y}[n,\ell]
    \frac{S_{\mathrm{SRS}}^*[n,\ell]}{\sqrt{\mathcal{P}_s}}\\
    &=
    \sqrt{\mathcal{P}_s}\,H[n,\ell]+W[n,\ell].
\end{aligned}
\end{equation}
The descrambled noise
$W[n,\ell]=\widetilde{W}[n,\ell]
S_{\mathrm{SRS}}^*[n,\ell]/\sqrt{\mathcal{P}_s}$
therefore retains the receiver-noise distribution
$\mathcal{CN}(0,\sigma_w^2)$. Within one SRS burst, the block-fading
approximation gives $H[n,\ell]\approx H[n]$~\cite[ch.~8]{sesia2011lte}.

The occupied SRS tones are embedded in an $N$-point frequency grid and
transformed using a unitary IDFT:
\begin{equation}
\label{eq:supp-delay-domain-channel}
\begin{aligned}
    y[m]
    &=
    \sqrt{\mathcal{P}_s}\,h[m]+w_t[m],\\
    h[m]
    &=
    \frac{1}{\sqrt{N}}\sum_{n=0}^{N-1}
    H[n]\exp\!\left(j2\pi\frac{nm}{N}\right).
\end{aligned}
\end{equation}
Here $h[m]$ is the sampled delay-domain channel coefficient within one burst,
under the block-fading assumption; restoring the burst index $s$ gives the
$y[m,s]$ and $h[m,s]$ used in~\eqref{eq:delay-domain-channel} and the slow-time
processing below. The unitary scaling preserves the noise variance, so
$w_t[m]\sim\mathcal{CN}(0,\sigma_w^2)$. For subcarrier spacing $\Delta_f$, the
delay-bin spacing is $T_s=1/(N\Delta_f)$ and the center of bin $m$ is
$\tau_m=mT_s$. The coefficient $h[m]$ aggregates the MPCs falling into, or
leaking into, the $m$th bin. For a non-contiguous or comb-like SRS allocation,
it is interpreted as the sampled response of the effective SRS waveform on this
delay grid, with off-grid leakage absorbed into the effective disturbance.
The unitary scaling preserves the \emph{total} noise variance, but a sparse comb
populates only $M_{\mathrm{SRS}}$ of the $N$ grid tones, so the transform of the
zero-filled spectrum yields delay-bin noise that is correlated across bins and
no longer strictly white. We adopt $\sigma_w^2$ as the post-processing
per-resolved-bin variance at the selected peaks $(m_0,m_k)$; the residual
inter-bin correlation, like the off-grid leakage, is folded into the effective
disturbance and is consistent with the genie-aided separability assumption used
throughout.

\subsection{Physical Path Coefficients and Doppler Projection}

Within burst $s$, the sampled channel coefficient can be written as a sum of
MPCs,
\begin{equation}
\label{eq:supp-h-bin-agg}
    h[m,s]
    =
    \sum_{u=0}^{M_m-1}
    a_{m,u}[s]\exp\!\left(-j\frac{2\pi}{\lambda}d_{m,u}\right),
\end{equation}
where $M_m$ is the number of MPCs contributing to bin $m$, $a_{m,u}[s]$ is the
complex attenuation, and $d_{m,u}$ is the total path length. For an MPC
composed of $L_{m,u}$ geometric segments, we use the attenuation model
\begin{equation}
\label{eq:supp-multibounce-attenuation}
    a_{m,u}
    =
    C_{m,u}\lambda
    \prod_{r=1}^{L_{m,u}}\frac{1}{4\pi d_{m,u,r}},
\end{equation}
where $C_{m,u}$ collects reflection, scattering, material, polarization, and
other wavelength-dependent effects not represented by the explicit distance
product.

For BD$_k$, the bistatic backscatter coefficient depends on its reflection
state $x^{(k)}[s]$:
\begin{equation}
\label{eq:supp-bd-channel}
\begin{aligned}
    a_{m,k}^{\mathrm{BD}}[s]
    &=
    \left(C^{(\mathrm{s})}
    +C^{(\mathrm{a})}x^{(k)}[s]\right)
    \widetilde{C}_{m,k}\lambda^2\\
    &\quad\times
    \prod_{r=1}^{L_{m,k}}\frac{1}{4\pi d_{m,k,r}},
\end{aligned}
\end{equation}
where $C^{(\mathrm{s})}$ and $C^{(\mathrm{a})}$ are the structural- and
antenna-mode coefficients, respectively, and $\widetilde{C}_{m,k}$ collects
the remaining interaction losses
~\cite{kimionis2013bistaticBackscatter,griffin2009completeLinkBudgets}. The
structural mode is independent of $x^{(k)}[s]$ and therefore remains in the
quasi-static environmental component, while the antenna mode is shifted to the
Doppler bin selected by the known switching sequence. In the notation of the
slow-time decomposition~\eqref{eq:slow-time-channel}, the antenna-mode term is
\begin{equation}
\label{eq:supp-hk-bd}
    h_k^{\mathrm{BD}}[m]
    =
    C^{(\mathrm{a})}\widetilde{C}_{m,k}\lambda^2
    \prod_{r=1}^{L_{m,k}}\frac{1}{4\pi d_{m,k,r}}
    \,e^{-j\frac{2\pi}{\lambda}d_{m,k}},
\end{equation}
multiplied by $x^{(k)}[s]$, whereas the BD structural mode and the direct NLOS
component together form $h_{\mathrm{env}}[m]$.

The unitary slow-time DFT of the switching sequence is
\begin{equation}
\label{eq:supp-doppler-spectrum}
    X^{(k)}[q]
    \triangleq
    \frac{1}{\sqrt{N_s}}\sum_{s=0}^{N_s-1}
    x^{(k)}[s]e^{-j2\pi sq/N_s}.
\end{equation}
For the ideal bin-aligned tone
$x^{(k)}[s]=e^{j2\pi s q_k/N_s}$,
$X^{(k)}[q_k]=\sqrt{N_s}$. The quasi-static direct NLOS component has the
corresponding zero-Doppler factor
$N_s^{-1/2}\sum_{s=0}^{N_s-1}1=\sqrt{N_s}$.

Under the single-dominant-MPC geometry of the main manuscript, the direct NLOS
path contains the UE-to-scatterer segment $d$ and the scatterer-to-BS segment
$d_{\mathrm{SB}}$. After selecting its delay--Doppler bin and removing the
known delay-bin-center phase, its measurement is
\begin{equation}
\label{eq:supp-direct-measurement}
    z_0
    =
    g_0
    \frac{e^{-j2\pi d/\lambda}}{d}
    +w_0,
\end{equation}
with effective gain
\begin{equation}
\label{eq:supp-direct-gain}
    g_0
    \approx
    \sqrt{N_s\mathcal{P}_s}\,C_0
    \frac{\lambda}{(4\pi)^2d_{\mathrm{SB}}}
    e^{-j2\pi d_{\mathrm{SB}}/\lambda}.
\end{equation}
Here $C_0$ absorbs the dominant-scatterer response and RF-chain factors not
represented by the explicit ranges.

The BD$_k$-assisted path contains the UE-to-BD distance
$d_{\mathrm{UE}}^{(k)}$, the BD-to-scatterer distance $d^{(k)}$, and the common
scatterer-to-BS segment $d_{\mathrm{SB}}$. Projection onto $q_k$ removes the
known sequence $x^{(k)}[s]$ from the extracted measurement and contributes the
known factor $X^{(k)}[q_k]$. The resulting measurement is
\begin{equation}
\label{eq:supp-bd-measurement}
    z_k
    =
    g_k e^{j\phi_k}
    \frac{
    e^{-j\frac{2\pi}{\lambda}
    (d_{\mathrm{UE}}^{(k)}+d^{(k)})}}
    {d_{\mathrm{UE}}^{(k)}d^{(k)}}
    +w_k,
\end{equation}
where $\phi_k$ is the residual phase not captured by the nominal switching
sequence, and
\begin{equation}
\label{eq:supp-bd-gain}
    g_k
    \approx
    \sqrt{\mathcal{P}_s}\,X^{(k)}[q_k]
    C^{(\mathrm{a})}\widetilde{C}_k
    \frac{\lambda^2}{(4\pi)^3d_{\mathrm{SB}}}
    e^{-j2\pi d_{\mathrm{SB}}/\lambda}.
\end{equation}
For the ideal bin-aligned tone, $X^{(k)}[q_k]=\sqrt{N_s}$, so $g_0$ in
\eqref{eq:supp-direct-gain} and $g_k$ in~\eqref{eq:supp-bd-gain} carry the same
$\sqrt{N_s\mathcal{P}_s}$ coherent-integration factor. Thus $x^{(k)}[s]$ is
absent from the compact measurement model because its slow-time DFT has already
been evaluated at $q_k$, with the resulting coefficient $X^{(k)}[q_k]$ included
in $g_k$. If the BDs are hardware-similar and use switching sequences with equal
projection magnitudes, the residual ratios $g_k/g_0$ can be approximated by the
common relative coefficient $\gamma$ used in~\eqref{eq:gain-fixed-approx}.

\section{Fisher Information for Ranging using Delay Estimation}
\label{appendix:range-delay-fi}

This appendix derives the delay-induced Fisher information term used in
\eqref{eq:delay-term}. Its purpose is to isolate the contribution of waveform
bandwidth to range estimation, so that the range-domain EFIMs in
Section~\ref{sec:crb-results} can include both narrowband channel-variation
terms and wideband delay information.

We work with a complex baseband representation in which the known sounding
waveform is denoted by $s(t)$ and the received waveform by $y(t;\tau)$. The
carrier phase rotation is absorbed into the complex channel coefficient
$a(\tau)$, while $B_{\mathrm{rms}}$ denotes the RMS bandwidth of the baseband
sounding waveform rather than the carrier frequency.

We keep the unknown complex channel coefficient $a(\tau)$ distinct from the known
sounding waveform. Let
\begin{equation*}
    s(t-\tau)=\sum_{k\in\mathcal{K}_{\mathrm{SRS}}} S[k]\,p(t-\tau)
    \exp\!\left(j2\pi f_k (t-\tau)\right)
\end{equation*}
be the \emph{gain-free} baseband waveform, with per-tone symbols $S[k]$, a
unit-energy, differentiable pulse shape $p(t)$, and baseband frequencies
$f_k=(k-k_{\mathrm{ref}})\Delta_f$, where $\mathcal{K}_{\mathrm{SRS}}$ collects
the active SRS subcarriers and $\Delta_f$ is the subcarrier spacing. The received
signal for a single resolved path is
\begin{equation*}
    y(t;\tau)=a(\tau)\,s(t-\tau)+w(t),
\end{equation*}
where $a(\tau)$ is the complex channel coefficient (which also absorbs the
carrier-phase rotation $\exp(-j2\pi f_c\tau)$) and $w(t)$ is white noise.
For the finite-energy interpretation used below, the observation interval is
assumed to contain the support of the effective waveform, so the boundary terms
vanish. A periodic OFDM useful-symbol model gives the analogous result through a
Fourier-series sum with cyclic boundaries. In either interpretation,
uncompensated discontinuous rectangular edges are excluded from the delay
derivative; practical transmit and receive filtering is included in the
effective pulse $p(t)$.

Differentiating the noiseless mean with respect to $\tau$,
\begin{equation*}
    \frac{\partial}{\partial\tau}\bigl[a(\tau)s(t-\tau)\bigr]
    =\frac{\partial a(\tau)}{\partial\tau}\,s(t-\tau)
    -a(\tau)\frac{\partial s(t-\tau)}{\partial t}.
\end{equation*}
The gain $a(\tau)$ appears exactly once in each term. Averaging the squared
magnitude over the useful symbol interval yields the normalized delay-sensitivity
energy $\mathcal{G}_{\tau}$. The cross-term is proportional to
\begin{equation*}
    \int_{0}^{T_{\mathrm{sym}}}
    s^{*}(t-\tau)\frac{\partial s(t-\tau)}{\partial t}\,\mathrm{d}t
    =
    j2\pi\bar{f}\int_{-\infty}^{\infty}|S(f)|^{2}\,\mathrm{d}f,
\end{equation*}
where the equality follows from Parseval's identity under the boundary
conditions stated above. It vanishes for the centered baseband spectrum, whose
spectral centroid is $\bar{f}=0$. Hence the two contributions separate as
\begin{equation*}
    \label{eq:Gamma-tau-derivation}
\begin{aligned}
    \mathcal{G}_{\tau}
    & = \frac{1}{T_{\mathrm{sym}}} \int_{0}^{T_{\mathrm{sym}}}
      \left|
      \frac{\partial a(\tau)}{\partial \tau}s(t-\tau)
      -a(\tau)\frac{\partial s(t-\tau)}{\partial t}
      \right|^2 \mathrm{d}t \\
    &= \frac{1}{T_{\mathrm{sym}}}\left|\frac{\partial a(\tau)}{\partial \tau}\right|^2 \int_{0}^{T_{\mathrm{sym}}} |s(t-\tau)|^2 \mathrm{d}t \\
    &\quad + \frac{1}{T_{\mathrm{sym}}} |a(\tau)|^2
    \int_{0}^{T_{\mathrm{sym}}}
    \left|\frac{\partial s(t-\tau)}{\partial t}\right|^2 \mathrm{d}t \\
    &= \underbrace{\left|\frac{\partial a(\tau)}{\partial \tau}\right|^2 \mathcal{P}_s}_{\text{Gain Variation}} + \underbrace{|a(\tau)|^2 (2\pi \mathrm{B}_{\mathrm{rms}})^2 \mathcal{P}_s}_{\text{Waveform Bandwidth}},
\end{aligned}
\end{equation*}
The first term accounts for the sensitivity of the channel coefficient itself
to delay changes. This is the coefficient-gradient information represented by
the range derivative of the compact-model mean in
Section~\ref{sec:crb-results}. The second term is the waveform-bandwidth
contribution, which is the term appearing in classical range-estimation
bounds~\cite{wang2009cramerRao}. It is driven by the average symbol power
$\mathcal{P}_s$ and the RMS bandwidth $\mathrm{B}_{\mathrm{rms}}$, defined as
\begin{equation}
\label{eq:Ps-Brms}
\begin{aligned}
    \mathcal{P}_s &\triangleq \frac{1}{T_{\mathrm{sym}}}\int_{0}^{T_{\mathrm{sym}}} |s(t)|^2 \mathrm{d}t, \\
    \mathrm{B}_{\mathrm{rms}}^2 &\triangleq \frac{\int_{-\infty}^{\infty} f^2 |S(f)|^2 \mathrm{d}f}{\int_{-\infty}^{\infty} |S(f)|^2 \mathrm{d}f},
\end{aligned}
\end{equation}
where $S(f)$ is the Fourier transform of the gain-free waveform $s$ (centered at
0~Hz). The received bin power is $|a(\tau)|^2\mathcal{P}_s$, which together with
the noise variance defines the per-path SNR used in the main text.

Finally, we translate the delay sensitivity to the physical range $d$.
The range is directly proportional to the time delay $\tau$ via the propagation
speed $c$, such that $d = c\tau$. Consequently, the normalized range-sensitivity
energy $\mathcal{G}_d$ scales with the inverse square of the propagation speed,
$
    \mathcal{G}_d \triangleq \frac{1}{c^2} \mathcal{G}_{\tau}.
$
Considering the bandwidth-dependent component of this sensitivity and using the
wave relation $c = \lambda f_c$, the effective
range resolution metric is given by
\begin{equation}
    \label{eq:range-sensitivity-delay}
    \mathcal{G}_d = \left|\frac{\partial a(d)}{\partial d}\right|^2 \mathcal{P}_s +
                |a(d)|^2 \left(\frac{2\pi}{\lambda} \frac{\mathrm{B}_{\mathrm{rms}}}{f_c}\right)^2 \mathcal{P}_s.
\end{equation}
This formulation demonstrates that for a fixed signal-to-noise ratio, the range
estimation precision is governed by the normalized bandwidth
$\mathrm{B}_{\mathrm{rms}}/f_c$ scaled by the wavenumber factor $2\pi/\lambda$.

In the Gaussian measurement model~\eqref{eq:mean-covariance-fixed-g} of Section~\ref{sec:crb-results}, the
deterministic waveform sensitivity is multiplied by the noise factor
$2/\sigma_w^2$. Equivalently, for the $i$th extracted path, the bandwidth-only
range information is
\begin{equation}
    \mathcal{I}^{(\mathrm{bw})}
    =
    \frac{2|{\mu}_i|^2}{\sigma_w^2}
    \left(\frac{2\pi}{\lambda} \frac{\mathrm{B}_{\mathrm{rms}}}{f_c}\right)^2
    =
    2\,\mathrm{SNR}_i\,\beta_B,
    \label{eq:fim-bw}
\end{equation}
where $\mathrm{SNR}_i\triangleq|{\mu}_i|^2/\sigma_w^2$ and
$\beta_B\triangleq(2\pi B_{\mathrm{rms}}/c)^2$. This is the scalar bandwidth
term added in~\eqref{eq:delay-term}; for a multi-segment path, the same scalar
multiplies the all-ones block associated with the segment lengths that form the
total path length.

\subsection{Example: 5G NR Sounding Reference Signal (SRS)}

In 5G New Radio (NR), the Sounding Reference Signal (SRS) is defined
in~\cite{threeGpp2023nrPhysicalChannels}. The SRS is designed as a ``comb'' structure in the
frequency domain, which allows the RMS bandwidth to be calculated in closed
form.

The RMS bandwidth is determined by the subcarrier spacing $\Delta_f$ and the
spread of the active subcarriers. For a discrete spectrum, the integral in
Eq.~\eqref{eq:Ps-Brms} is replaced by a summation:
\begin{equation}
\begin{aligned}
    B_{\mathrm{rms}}^2
    &= \Delta_f^2\,
      \frac{\sum_{k\in\mathcal{K}_{\mathrm{SRS}}} |S[k]|^2 (k-\bar{k})^2}{
            \sum_{k\in\mathcal{K}_{\mathrm{SRS}}} |S[k]|^2}, \\
    \bar{k}
    &= \frac{\sum_{k\in\mathcal{K}_{\mathrm{SRS}}} |S[k]|^2 k}{
            \sum_{k\in\mathcal{K}_{\mathrm{SRS}}} |S[k]|^2}.
\end{aligned}
\end{equation}
According to TS 38.211~\cite{threeGpp2023nrPhysicalChannels}, the SRS subcarriers are located at
indices $k = k_0 + p K_{\mathrm{TC}}$ for $p=0,\ldots,M_{\mathrm{SRS}}-1$,
where $K_{\mathrm{TC}} \in \{2, 4, 8\}$ is the transmission comb size, and
$M_{\mathrm{SRS}}$ denotes the total number of active subcarriers allocated to
the SRS. Furthermore, since Zadoff-Chu sequences are used, the sequence $S[k]$
has constant magnitude $|S[k]|^2 = \text{const.}$ across these active tones.

Under these conditions, the RMS bandwidth reduces to the variance of a uniform comb:
\begin{equation}
    B_{\mathrm{rms}}^2
    = \Delta_f^2\,K_{\mathrm{TC}}^2\,\frac{M_{\mathrm{SRS}}^2-1}{12}.
\end{equation}
This expression quantifies the fundamental resolution limit of the SRS
configuration. It highlights that resolution improves linearly with the comb
spacing $K_{\mathrm{TC}}$ (due to wider spectral spread for the same number of
tones) and the total number of allocated subcarriers $M_{\mathrm{SRS}}$.


\section{Proof of EFIM Results}
\label{app:efim-proofs}

This appendix proves the range-domain EFIM results stated in
Section~\ref{sec:crb-results}. We use the FIM expression
in~\eqref{eq:fim-formula}, include the waveform delay contribution
in~\eqref{eq:delay-term}, and then apply path-specific derivatives together
with Schur-complement elimination of the corresponding nuisance parameters.
For compactness, we use the SNR definitions in~\eqref{eq:snr-def} and define
\begin{equation}
    \mathbf{D}_{\mathrm{SNR}}
    \triangleq
    \diag(\mathrm{SNR}_1,\ldots,\mathrm{SNR}_K).
\end{equation}

The range-parameter derivatives used throughout the appendix are
\begin{subequations}
\label{eq:app-r-derivatives}
\begin{align}
    \frac{\partial \rho_0}{\partial d}
    &=
    -\!\left(\frac{1}{d}+j\frac{2\pi}{\lambda}\right)\rho_0,\\
    \frac{\partial \rho_k}{\partial d_{\mathrm{UE}}^{(k)}}
    &=
    -\!\left(\frac{1}{d_{\mathrm{UE}}^{(k)}}+j\frac{2\pi}{\lambda}\right)\rho_k,\\
    \frac{\partial \rho_k}{\partial d^{(k)}}
    &=
    -\!\left(\frac{1}{d^{(k)}}+j\frac{2\pi}{\lambda}\right)\rho_k.
\end{align}
\end{subequations}
The nuisance-parameter derivatives are obtained from
$\boldsymbol{\mu}=g\boldsymbol{\Gamma}\boldsymbol{r}$ as
\begin{subequations}
\label{eq:app-nuisance-derivatives}
\begin{align}
    \frac{\partial\boldsymbol{\mu}}{\partial g}
    &=
    \boldsymbol{\Gamma}\boldsymbol{r},\\
    \frac{\partial\boldsymbol{\mu}}{\partial\gamma}
    &=
    g(\mathbf{I}_{K+1}-\mathbf{e}_1\mathbf{e}_1^\top)\boldsymbol{r},\\
    \frac{\partial\boldsymbol{\mu}}{\partial\phi_k}
    &=
    j\gamma g r_k\mathbf{e}_{k+1}.
\end{align}
\end{subequations}
For complex parameters, the Wirtinger rule gives
$\partial\boldsymbol{\mu}/\partial\Re\{g\}=\partial\boldsymbol{\mu}/\partial g$
and
$\partial\boldsymbol{\mu}/\partial\Im\{g\}=j\,\partial\boldsymbol{\mu}/\partial g$
(similarly for $\gamma$).

\subsection{Proof of Proposition~\ref{prop:calibrated-efim}}

When $g$, $\gamma$, and $\boldsymbol{\phi}$ are known, only the
range-parameter derivatives in~\eqref{eq:app-r-derivatives} are needed.
Because $[d_{\mathrm{UE}}^{(k)},d^{(k)}]$ affects only $\rho_k$, the FIM is
block diagonal. Substituting~\eqref{eq:app-r-derivatives}
into~\eqref{eq:fim-formula} and adding~\eqref{eq:delay-term} gives
the blocks in~\eqref{eq:fim-blocks-cal}, proving
Proposition~\ref{prop:calibrated-efim}.

\subsection{Proof of Proposition~\ref{prop:unknown-phase-efim}}

The phase nuisance FIM block is
\begin{equation}
    \boldsymbol{\mathcal{I}}_{\boldsymbol{\phi}\boldsymbol{\phi}}
    =
    2\mathbf{D}_{\mathrm{SNR}}.
\end{equation}
The cross-term between $\phi_k$ and each BD distance parameter $\eta\in
\{d_{\mathrm{UE}}^{(k)},d^{(k)}\}$ evaluates to
$-2\mathrm{SNR}_k(2\pi/\lambda)$. In matrix form,
\begin{equation}
    \boldsymbol{\mathcal{I}}_{\boldsymbol{\phi}\boldsymbol{d}}
    =
    -2\frac{2\pi}{\lambda}\,
    \mathbf{D}_{\mathrm{SNR}}\mathbf{S},
\end{equation}
where $\mathbf{S}\in\{0,1\}^{K\times(2K+1)}$ is the path-segment selection
matrix
\begin{equation}
    \mathbf{S}
    \triangleq
    \bigl[\mathbf{0}_{K\times1},\,\mathbf{I}_K\otimes\mathbf{1}_2^\top\bigr],
    \qquad
    \mathbf{1}_2^\top=[1,1],
\end{equation}
where $\otimes$ denotes the Kronecker product.
Thus the $k$th row of $\mathbf{S}$ selects the two BD-assisted range
parameters $(d_{\mathrm{UE}}^{(k)},d^{(k)})$ and leaves the direct range $d$
unselected. The Schur loss is
\begin{equation}
\boldsymbol{\mathcal{I}}_{\boldsymbol{d}\boldsymbol{\phi}}
\boldsymbol{\mathcal{I}}_{\boldsymbol{\phi}\boldsymbol{\phi}}^{-1}
\boldsymbol{\mathcal{I}}_{\boldsymbol{\phi}\boldsymbol{d}} 
= 
2 \left(\frac{2\pi}{\lambda}\right)^2
\mathbf{S}^\top\mathbf{D}_{\mathrm{SNR}}\mathbf{S}.
\end{equation}
It is block diagonal with $2\times2$ BD blocks
$2\mathrm{SNR}_k(2\pi/\lambda)^2\mathbf{1}_2\mathbf{1}_2^\top$, which exactly
cancel the carrier-phase term of every BD block in~\eqref{eq:fim-blocks-cal},
leaving~\eqref{eq:I-k-phi}.
This proves Proposition~\ref{prop:unknown-phase-efim}.

\subsection{Proof of Proposition~\ref{prop:unknown-gain-bd-coefficient-efim}}

Let
$\boldsymbol{\alpha}=[\Re\{g\},\Im\{g\},\Re\{\gamma\},\Im\{\gamma\}]^\top$.
Define
$R_\gamma^2\triangleq|r_0|^2+|\gamma|^2\|\mathbf{r}_{\mathrm{BD}}\|^2$.
The joint nuisance FIM is
\begin{equation}
    \boldsymbol{\mathcal{I}}_{\boldsymbol{\alpha}\boldsymbol{\alpha}}
    =
    \begin{bmatrix}
    \boldsymbol{\mathcal{I}}_{gg} & \boldsymbol{\mathcal{I}}_{g\gamma}\\
    \boldsymbol{\mathcal{I}}_{\gamma g} & \boldsymbol{\mathcal{I}}_{\gamma\gamma}
    \end{bmatrix}.
\end{equation}
Using the block inverse, its inverse is
\begin{equation}
    \boldsymbol{\mathcal{I}}_{\boldsymbol{\alpha}\boldsymbol{\alpha}}^{-1}
    =
    \frac{\sigma_w^2}{2|g|^2|r_0|^2}
    \begin{bmatrix}
    |g|^2\mathbf{I}_2 & -\mathbf{R}_{g\gamma}\\
    -\mathbf{R}_{g\gamma}^{\top}
    & \left(
    \frac{|r_0|^2}{\|\mathbf{r}_{\mathrm{BD}}\|^2}+|\gamma|^2
    \right)\mathbf{I}_2
    \end{bmatrix}\!,
\end{equation}
with
\begin{subequations}
\begin{align}
\boldsymbol{\mathcal{I}}_{gg}
&=
\frac{2}{\sigma_w^2}R_\gamma^2\mathbf{I}_2,\\
\boldsymbol{\mathcal{I}}_{\gamma\gamma}
&=
\frac{2|g|^2}{\sigma_w^2}
\|\mathbf{r}_{\mathrm{BD}}\|^2\mathbf{I}_2,\\
\mathbf{R}_{g\gamma}
&=
\begin{bmatrix}
\Re\{g\gamma^*\} & -\Im\{g\gamma^*\}\\
\Im\{g\gamma^*\} & \Re\{g\gamma^*\}
\end{bmatrix}.
\end{align}
\end{subequations}

The cross-terms between $\boldsymbol{\alpha}$ and the distance parameters
are computed from~\eqref{eq:app-nuisance-derivatives}.
Substituting into the Schur complement yields Schur-loss blocks
\begin{subequations}
\begin{align}
    \mathcal{I}_{\mathrm{loss},d}^{(g,\gamma)}
    &=
    2\mathrm{SNR}_0
    \!\left[
    \frac{1}{d^2}+\!\left(\frac{2\pi}{\lambda}\right)^{\!2}
    \right]\!,\\
    \boldsymbol{\mathcal{I}}_{\mathrm{loss},k\ell}^{(g,\gamma)}
    &=
    2\mathrm{SNR}_k
    \frac{|r_\ell|^2}{\|\mathbf{r}_{\mathrm{BD}}\|^2}
    \!\left[
    \mathbf{v}_k\mathbf{v}_\ell^\top
    +\!\left(\frac{2\pi}{\lambda}\right)^{\!2}\!\mathbf{1}_2\mathbf{1}_2^\top
    \right]\!.
\end{align}
\end{subequations}
Subtracting from the calibrated FIM gives
\eqref{eq:I-d-gain}--\eqref{eq:I-kl-gain}, proving
Proposition~\ref{prop:unknown-gain-bd-coefficient-efim}.

\subsection{Proof of Corollary~\ref{cor:unknown-common-gain}}

Only the common gain $g$ is eliminated. The narrowband distance derivatives can
be written as the
calibrated path coefficients multiplied by
$d^{-1}+j2\pi/\lambda$ for the direct path and by
$\mathbf{v}_k+j(2\pi/\lambda)\mathbf{1}_2$ for the $k$th BD-assisted path, up
to signs that vanish in the real Fisher products. The Schur loss induced by
eliminating $g$ is therefore the projection of these derivatives onto the
one-dimensional complex subspace spanned by $\boldsymbol{\Gamma}\boldsymbol{r}$.
The direct-path, direct--BD, and BD--BD Schur-loss blocks are
\begin{subequations}
\begin{align}
    \mathcal{I}_{\mathrm{loss},d}^{(g)}
    &=
    2\mathrm{SNR}_0
    \frac{|r_0|^2}{\|\boldsymbol{\Gamma}\boldsymbol{r}\|^2}
    \left[
    \frac{1}{d^2}+\left(\frac{2\pi}{\lambda}\right)^2
    \right],\\
    \boldsymbol{\mathcal{I}}_{\mathrm{loss},d k}^{(g)}
    &=
    2\mathrm{SNR}_0
    \frac{|\gamma|^2|r_k|^2}{\|\boldsymbol{\Gamma}\boldsymbol{r}\|^2}
    \left[
    \frac{1}{d}\mathbf{v}_k^\top
    +\left(\frac{2\pi}{\lambda}\right)^2\mathbf{1}_2^\top
    \right],\\
    \boldsymbol{\mathcal{I}}_{\mathrm{loss},k\ell}^{(g)}
    &=
    2\mathrm{SNR}_k
    \frac{|\gamma|^2|r_\ell|^2}
    {\|\boldsymbol{\Gamma}\boldsymbol{r}\|^2}
    \left[
    \mathbf{v}_k\mathbf{v}_\ell^\top
    +\left(\frac{2\pi}{\lambda}\right)^2
    \mathbf{1}_2\mathbf{1}_2^\top
    \right].
\end{align}
\end{subequations}
Subtracting these losses from the calibrated block-diagonal EFIM, while leaving
the wideband delay terms unchanged, gives \eqref{eq:I-d-g-only} and
\eqref{eq:I-dk-g-only}, together with the BD-assisted blocks
\eqref{eq:I-kk-g-only} and~\eqref{eq:I-kl-g-only}. Unlike the joint
unknown-$\{g,\gamma\}$ case, the BD--BD loss is normalized by
$\|\boldsymbol{\Gamma}\boldsymbol{r}\|^2
=|r_0|^2+|\gamma|^2\|\mathbf{r}_{\mathrm{BD}}\|^2$. The direct NLOS component
therefore helps constrain $g$ when $\gamma$ is known.

\subsection{Proof of Corollary~\ref{cor:individual-g-gamma}}

Here $g$ is known and only the complex scalar $\gamma$ is eliminated. Then
\begin{equation}
    \boldsymbol{\mathcal{I}}_{\gamma\gamma}^{-1}
    =
    \frac{\sigma_w^2}{2|g|^2\|\mathbf{r}_{\mathrm{BD}}\|^2}\mathbf{I}_2,
\end{equation}
and the direct-path cross block is zero because $\gamma$ does not enter
$z_0$. The resulting BD-to-BD Schur loss is again
$\boldsymbol{\mathcal{I}}_{\mathrm{loss},k\ell}^{(g,\gamma)}$ from the proof
of Proposition~\ref{prop:unknown-gain-bd-coefficient-efim}. Hence the direct
block remains~\eqref{eq:I-d-cal}, while the BD-assisted blocks are those
in~\eqref{eq:I-kk-gain} and~\eqref{eq:I-kl-gain}.

\subsection{Proof of Corollary~\ref{cor:fully-unknown-efim}}

When $g$, $\gamma$, and $\boldsymbol{\phi}$ are all unknown, the phase of
$\gamma$ is absorbed into the per-BD phases,
$\tilde{\phi}_k\triangleq\phi_k+\angle\gamma$. Applying the phase Schur
complement first removes the carrier-phase term of each BD block, as in
Proposition~\ref{prop:unknown-phase-efim}. The remaining dependence on the
relative BD coefficient is through the real amplitude $|\gamma|$. Applying the
joint $\{g,|\gamma|\}$ Schur complement then removes the direct path's narrowband
terms and subtracts the common-amplitude loss
\begin{equation}
\label{eq:full-common-amplitude-loss}
    \boldsymbol{\mathcal{L}}_{k,\ell}^{(g,|\gamma|)}
    =
    2\mathrm{SNR}_k
    \frac{|r_\ell|^2}{\|\mathbf{r}_{\mathrm{BD}}\|^2}
    \mathbf{v}_k\mathbf{v}_\ell^\top
\end{equation}
from each BD block pair. Subtracting
\eqref{eq:full-common-amplitude-loss} from the phase-eliminated EFIM in
\eqref{eq:I-k-phi} yields the diagonal and off-diagonal blocks in
\eqref{eq:I-kk-full} and~\eqref{eq:I-kl-full}; the direct path retains only
the bandwidth term in~\eqref{eq:I-d-full}. This proves
Corollary~\ref{cor:fully-unknown-efim}.


\bibliographystyle{IEEEtran}
\bibliography{refs}

\end{document}